\documentclass[pre,preprint,superscriptaddress,amsmath,amssymb,floatfix]{revtex4}
\usepackage{graphicx}
\usepackage{epsfig}
\usepackage{dcolumn}
\usepackage{bm}
\usepackage{psfrag}
\usepackage{color}

\newcommand{\Qvec}{\mathbf{Q}}
\newcommand{\Qij}{Q_{ij}}

\newcommand{\dd}{\mathrm{d}}

\newcommand{\NC}{\mathrm{N}_\mathrm{C}}
\newcommand{\ND}{\mathrm{N}_\mathrm{D}}
\newcommand{\dir}{\mathbf{\hat{n}}}

\definecolor{forestgreen}{rgb}{0.133333,0.545098,0.133333}

\DeclareMathOperator{\Tr}{Tr}

\begin{document}

\bibliographystyle{apsrev}

\title{Dispersions of ellipsoidal particles in a nematic liquid crystal}
\author{Mykola Tasinkevych}
\email{miko@mf.mpg.de} \affiliation{Max-Planck-Institut f\"ur Intelligente Systeme, Heisenbergstr. 3, D-70569
Stuttgart, Germany} \affiliation{Institut f\"{u}r Theoretische Physik IV,
         Universit\"{a}t Stuttgart, Pfaffenwaldring 57,
         D-70569 Stuttgart, Germany}
\author{Fr\'{e}d\'{e}ric Mondiot}
\affiliation{Universit\'{e} de Bordeaux, CNRS, Centre de Recherche Paul Pascal, Avenue A. Schweitzer F-33600
Pessac, France}
\author{Olivier Mondain-Monval}
\affiliation{Universit\'{e} de Bordeaux, CNRS, Centre de Recherche Paul Pascal, Avenue A. Schweitzer F-33600
Pessac, France}
\author{Jean-Christophe Loudet}\email{loudet@crpp-bordeaux.cnrs.fr}
\affiliation{Universit\'{e} de Bordeaux, CNRS, Centre de Recherche Paul Pascal, Avenue A. Schweitzer F-33600
Pessac, France}
\date{\today}

\begin{abstract}
Colloidal particles dispersed in a partially ordered medium, such as a liquid crystal (LC) phase, disturb its
alignment and are subject to elastic forces. These forces are long-ranged, anisotropic and tunable through
temperature or external fields, making them a valuable asset to control colloidal assembly. The latter is very
sensitive to the particle geometry since it alters the interactions between the colloids. We here present a
detailed numerical analysis of the energetics of elongated objects, namely prolate ellipsoids, immersed in a
nematic host. The results, complemented with qualitative experiments, reveal novel LC configurations with
peculiar topological properties around the ellipsoids, depending on their aspect ratio and the boundary
conditions imposed on the nematic order parameter. The latter also determine the preferred orientation of
ellipsoids in the nematic field, because of elastic torques, as well as the morphology of particles aggregates.
\end{abstract}

%\pacs{68.03.Hj,68.18.Fg,68.18.Jk,68.37.Ps}
\maketitle

\section{Introduction}

\noindent Dispersions of colloidal particles in anisotropic solvents such as liquid crystals (LC) make up a new
class of composite materials that was discovered at the end of the last century
\cite{bury,tere,ruhw,poul97,poul98,lube,monda99,star99}. Since then, there has been a surge of activity in this
area covering both fundamental as well as more applied issues \cite{star01,muse}. In such systems the colloidal
inclusions experience new interactions -- the so-called elastic interactions -- that do not exist in more usual
colloidal dispersions made of isotropic solvents. These interactions are long ranged, anisotropic, and have an
elastic character originating from the distortions of the LC intrinsic order in the vicinity of the inclusions.
The amplitude and symmetry of these distortions, which cost elastic energy, are to a large extent controlled by
the boundary conditions at the colloid-LC interface. The latter are defined by the LC anchoring conditions, i.e.
the orientation and strength with which the LC molecules are bound to surfaces (liquid or solid). Boundaries
therefore play a crucial role in LC colloids for they control the way the particles interact with each other,
which has a direct impact on the physical properties of the system.

Most of the studies performed in the past 15 years focused on the properties of micrometer-sized spherical
colloids dispersed in a nematic LC phase. The topological defects, the director topology, the (anisotropic)
elastic interactions between inclusions and the influence of various boundary conditions as well as external
fields were all extensively characterized \cite{tere,ruhw,poul97,poul98,lube,monda99,star99,
star01,muse,gu,grol,skar_a,skar_b,loud01,fuku,rama,poul97b,smal05,taka,eska}. The long range nature of elastic forces
(inverse power laws) is responsible for the building of large-scale ordered assemblies (linear chains, 2D and 3D
crystals), while topological defects mainly ensure stability of the system
\cite{muse,loud00,ravn,ogny08,ogny09,nych}. Typical pair interaction energies for colloidal sizes $\sim\mu$m are
of the order $\sim 10^3\,k_BT\,$, meaning that the resulting particles' aggregates are insensitive to thermal
fluctuations and remain stable over time.

Today, new effects and properties are being unveiled such as the newly coined ``nematic braids'' defects
\cite{tkal11} or topological colloids \cite{senyuk:13,liu:13}. However, the dynamical aspects of such systems
still remain largely unexplored despite a few recent studies \cite{lavr,pish,lint}. Furthermore, the interest in
the field is further boosted by the recent progresses made in synthesizing new model particles varying in size,
shape, nature, structure (core-shell) and surface properties as reviewed in \cite{cham,saca10,saca11}.
Synergetic properties between the inclusions and the LC matrix are now being explored, with the long term
objective of designing materials with extraordinary properties, such as those expected for metamaterials for
instance \cite{liu11}.

Most of the foreseen applications require rather high particle volume fractions, i.e. $\sim 10\%$ or more.
However, it is still a puzzling challenge today to achieve homogeneous colloidal dispersions with tunable
interparticle separations at such high particle loads. Because of the very high interaction energy ($\sim
10^2-10^3\,k_BT\,$), particles aggregate massively leading to macroscopic phase separations in most cases. Note
however that spectacular rheological properties were recently discovered with a colloid-nematic LC composite
where the particle volume fraction exceeded $20\%\,$ \cite{wood}.

Tuning the LC anchoring properties at the particles surface is a worthy route to circumvent the outlined
problem: indeed, low anchoring strengths (and hence low surface energies) allow for a partial relaxation of
elastic deformations in the bulk, thereby decreasing the overall elastic energy of the system. With liquid
inclusions, anchoring properties may be readily altered due to adsorbed surfactants and polymers
\cite{poul98,lock}. This is however not the case with solid inclusions which require specific grafting
procedures of chemical groups at the particle surface. Use of photosensitive molecules may bring up new insights
though, as shown in \cite{prat}: the LC anchoring type can be dynamically and reversibly switched between normal
and tangential anchoring states.

The particle size is another parameter to play with in order to modulate the anchoring properties, as reported
previously \cite{volt,koen09a,toma,star01}. Particles smaller (larger) than the typical length $K/W\sim\mu$m,
where $K$ is the (average) nematic elastic constant ($\sim 10\,$pN) and $W$ the surface anchoring strength
($\sim 5\times10^{-6}\,\mathrm{J.m}^{-2}$) \cite{pgg,pier}, tend to favor weak (strong) anchoring conditions
with a low (high) cost in terms of elastic energy. Indeed, recent experiments with nanometer-sized inclusions
revealed that pair interaction energies of just a few $k_BT$ could be achieved in thermotropic nematics
\cite{koen09b}; a promising result in view of the long term applications mentioned above.

An alternative strategy to achieve high particle loads consists in playing with the particle shape. Mondiot
\textit{et al.} indeed showed that ellipsoidal inclusions with tangential anchoring conditions tend ``to go
unnoticed'' when dispersed in lyotropic nematics, provided they are elongated enough \cite{mondi}. At small
volume fractions ($\leq 0.01\%$), the micrometer-sized ellipsoids align their long axis parallel to the local
director and remain individually dispersed over long periods of time (several months). Similar observations were
made for bacteria and gold nanorods embedded in other lyotropic nematic phases \cite{smal08,liu10}. However, in
other experiments involving non-spherical inclusions in thermotropic nematics, aggregation occurs
\cite{tkal08,lapo09}.

In the present work, we offer a thorough numerical analysis of the energetics of ellipsoidal particles embedded
in a nematic phase, complemented with some experimental observations. The overall objective is to gain insights
on the possibility of using both particle geometry and surface anchoring properties (tangential versus
homeotropic) in order to achieve homogeneous LC dispersions with a large amount of particles. Both single and
pairs of ellipsoids are considered together with various anchoring types and strengths at the LC-colloid
interface. The results show that, for a given anchoring strength, the nature of the anchoring controls the
orientation of ellipsoids with respect to the far-field director. Varying the anchoring strength and the
particle aspect ratio reveals peculiar director configurations and topological defects in the vicinity of the
inclusions, leading to metastable states in the free energy landscape. Numerically computed pair interaction
energies emphasize a significant influence of the anchoring strength on the morphology of particles' aggregates.
The experimental observations qualitatively support the calculations. \\

\section{ Landau-de Gennes model}

\noindent Within the Landau-de Gennes (LdG) theory, nematic liquid crystals are characterised by a
traceless symmetric order-parameter tensor with components $\Qij$ and the corresponding LdG free energy
functional is \cite{pgg}
\begin{equation}
F = \int_{\Omega} (f_{b} + f_{el})\,\dd^3x + \int_{\partial\Omega} f_{s}\,\dd s\,,
\label{free_energy}
\end{equation}
with $f_{b}$ and $f_{el}$ the bulk and elastic free energy densities, given by
\begin{eqnarray}
 f_{b} &=& a \Tr \Qvec^2 - b \Tr
\Qvec^3 + c \left(\Tr \Qvec^2\right)^2 \label{bulk},\\
f_{el} &=& \frac{L_1}{2}\partial_k \Qij\partial_k \Qij + \frac{L_2}{2} \partial_j \Qij
\partial_k Q_{ik}\,,
\label{elastic}
\end{eqnarray}
where $a$ depends linearly on the temperature $T$ and is usualy written as $a = a_0 (T - T^*)$, with $a_0$ a
material dependent constant and $T^*$ the supercooling temperature of the isotropic phase. $b$ and $c$ are
positive (material dependent) constants, and $L_1$ and $L_2$ are phenomenological parameters which can be
related to the Frank-Oseen (FO) elastic constants. The first integral in Eq.~(\ref{free_energy}) is taken over
the 3D domain, $\Omega$, occupied by nematic, while the second integral is over the surfaces $\partial\Omega$
(in our case the surfaces of the colloidal particles) and accounts for non-rigid anchoring conditions.

{
Depending on the anchoring conditions, various expressions are at our disposal to model the surface free energy
density $f_s\,$.
In the case of homeotropic anchoring, we used the following simple quadratic function favouring monostable
nematic ordering ${\Qvec^s}$, i.e., with a well-defined director, scalar and biaxial order parameters}
\cite{Nobili.1992}:
\begin{equation}
f_s=W\left(\Qij-\Qij^s\right)^2\,, \label{Nobili}
\end{equation}
{where $W >0$ is the anchoring strength at the colloidal surface whereas $\Qij^s$ describes the preferred
surface nematic ordering defined by $\Qij^s= 3 Q_b\left(\nu_i \nu_j -\delta_{ij}/3 \right)/2\,$, where
$\bm{\nu}$ is the unit normal to the colloidal surface. $Q_b$ is the bulk value of the scalar order parameter
(see Supplementary
 Information (SI) for the definition). We model planar
degenerate anchoring at the colloidal surfaces by the following surface potential \cite{Fournier.2005}}

\begin{equation}
f_s=W_1\left(\tilde{Q}_{ij}-\tilde{Q}_{ij}^\perp\right)^2 + W_2\left[\tilde{Q}_{ij}^2 -
\Bigl(\frac{3Q_b}{2}\Bigr)^2 \right]^2\,, \label{Fournier}
\end{equation}
{ where $\tilde{Q}_{ij}=\Qij+Q_b\frac{\delta_{ij}}{2}$,
$\tilde{Q}_{ij}^\perp=\left(\delta_{ik}-\nu_i\nu_k\right)\tilde{Q}_{ij}\left(\delta_{lj}-\nu_l\nu_j\right)$.
 $W_1>0$ is the anchoring strength favoring tangential
orientation of the director, and $W_2>0$ ensures the existence of a minimum for the surface scalar order
parameter at $Q_b\,$. We assume for simplicity $W_1=W_2\equiv W$. We also introduce the dimensionless anchoring
strength $w=W Q_b^2R_0/K_2\,$, where $R_0$ is some typical length scale related  to the size of colloidal
particle and $K_2$ is the Frank elastic constant for twist deformations. Relations between the Frank elastic
constants $K_i$ and the phenomelogical parameters $L_1,L_2$ can be found in the SI together with additional
material concerning our numerical procedure. }

\section{Results and discussion}

\noindent
In this section, we provide a description of the results obtained through both numerical calculations and
qualitative observations. We begin with an isolated micrometer-sized prolate ellipsoid embedded in a nematic
domain prior to dealing with pairs of such particles. As aforementioned, we investigate the effects of particle
aspect ratio, strength and nature of boundary conditions at the LC-colloid interface.

\subsection{Single ellipsoid}

\subsubsection{Numerical computations}

\noindent The case of spheres with strong homeotropic anchoring conditions and a homogeneous nematic director in the far
field, was thoroughly analyzed in the literature. Very briefly, a dipolar configuration, consisting in a
point-like hedgehog defect close to the sphere, was both computed and evidenced experimentally
\cite{ruhw,poul97,poul98,lube,star99}. A quadrupolar configuration, in which a disclination line encircles the
sphere at the equator -- the so-called Saturn ring configuration -- offers an alternate geometry to the dipolar
state \cite{tere,ruhw,monda99,star01,gu}. Now, how is the director topology altered in cases of elongated
particles? We shall first consider a single ellipsoid with strong homeotropic anchoring conditions. \\

\begin{figure}
\centering
\includegraphics[width=0.9\textwidth]{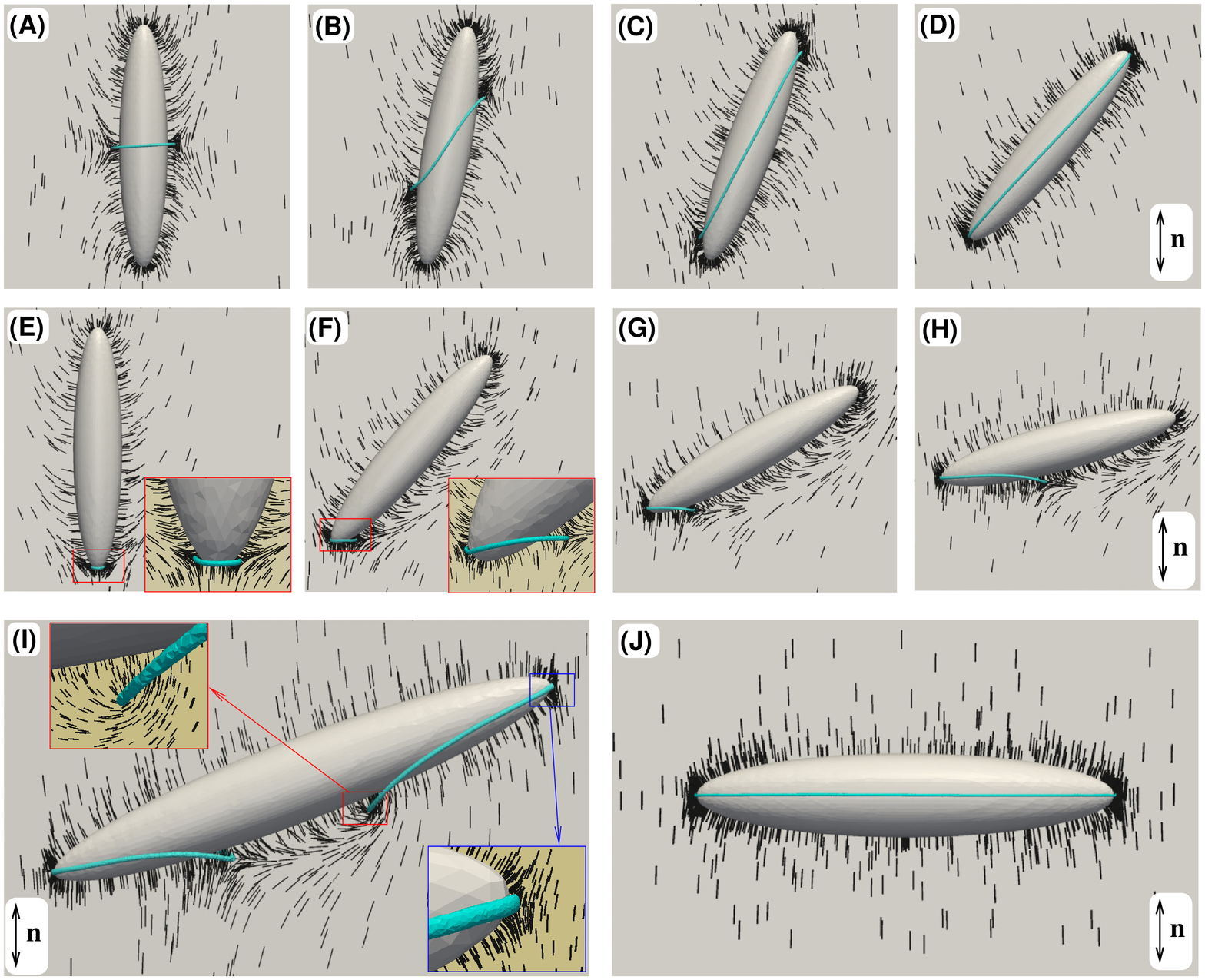}
\caption{Typical director configurations (black rods) around an ellipsoidal particle with homeotropic anchoring
($w = 37$). In  \textbf{(I)}, the aspect ratio $k = 7$, while in all other cases $k = 5$.  \textbf{(A)}-- \textbf{(D)}:
midplane ring configurations for $\theta = 0^\circ, 10^\circ, 20^\circ ,40^\circ$ respectively; $\theta$ is
an angle between the far-field director and the long axis of the ellipsoid.
 \textbf{(E)}--\textbf{(H)}: tip ring configurations for $\theta = 0^\circ, 40^\circ, 60^\circ ,75^\circ$
respectively. \textbf{(I)} shows metastable double ring configurations at $\theta = 70 ^\circ$.
{ Insets in \textbf{(E)},  \textbf{(F)} show zoomed-in views of  the director profiles
in the vicinity of the tip rings. The red and blue insets in \textbf{(I)} present detailed view of
 $1/2$ and $-1/2$ disclination profiles, respectively. }Isosurfaces of constant reduced scalar order parameter, $Q =0.6 Q_b\,$,
are shown in blue, where $Q_b$ is the bulk value of the scalar order parameter
(see the ``Methods'' section for the definition of $Q_b$, as well as other model
 parameters).} \label{1body_configs}
\end{figure}

 \noindent \textit{Strong homeotropic anchoring} -- Starting with various initial director configurations, we minimized the
Landau-de Gennes free energy (Eq.~(\ref{free_energy})) according to the procedure described in the ``Methods''
section. Figs.~\ref{1body_configs}A and E show two metastable configurations while Fig.~\ref{1body_configs}J
displays the most stable state. The former configuration (\ref{1body_configs}A) consists in a disclination line
that encircles the particle in its midplane when the ellipsoid long axis is parallel to $\dir\,$. This
topological defect is the analog of the Saturn ring defect encountered with spherical colloids and appears as a
consequence of the conservation of the total topological charge $Q$ in the nematic \cite{ruhw,lube,star99}. The
obtained director topology, which we will call ``midplane ring'' in the following, is similar to that previously
computed for rod-shaped objects \cite{andr,hung} and appear from an initial uniform director field.

 The other configuration shown in Fig.~\ref{1body_configs}E is a new director morphology which, as far
as we know, was not predicted by previous simulations on rod-like objects. It consists of a small disclination
ring located at one tip of the ellipsoid, which will be referred to as ``tip ring'' hereafter. Similarly to the
midplane ring above, the tip ring carries a topological point charge equal to $-1$ to ensure a zero global
topological charge. The tip ring shows up from hedgehog or non-equatorial ring initial configurations of the
director field. As we will see shortly, this topology has a lower energy than the midplane ring when the tilt
angle, $\theta$, is zero, i.e. the ellipsoid long axis is parallel to $\dir\,$.

\begin{figure}
\centering
\includegraphics[width=0.8\textwidth]{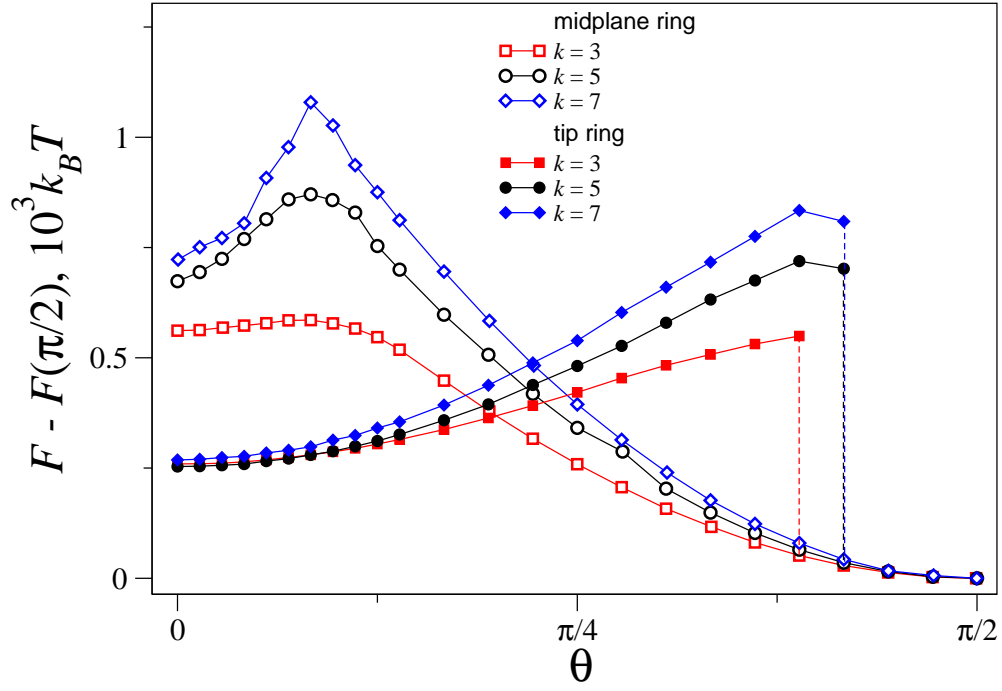}
\caption{Landau-de Gennes free energy, Eq.~(\ref{free_energy}), as a function of the angle $\theta$ between the
far-field director and the long axis of the ellipsoid for several values of the aspect ratio $k$ in
the case of strong homeotropic anchoring, $w = 37$ (see the ``Methods'' section for the
 definition of $w\,$, as well as other model parameters). Open (solid) symbols correspond to
the free energy branches of midplane ring (tip ring) configurations.} \label{en_vs_k}
\end{figure}

Unlike spheres, anisotropic particles will experience torques and hence adopt preferential orientations with
respect to $\dir\,$, as already reported \cite{mondi,smal08,liu10,tkal08,andr,hung,lapo04}. We therefore
computed the total free energy (Eq.~(\ref{free_energy})) as a function of $\theta$ for several aspect ratios $k$
and strong homeotropic conditions (dimensionless anchoring strength $w=37$; see the ``Methods'' section).
The results are displayed in Fig. \ref{en_vs_k}. Curves with open symbols correspond to midplane ring
configurations (Fig.~\ref{1body_configs}A-D), whereas curves with solid symbols refer to ``tip ring'' ones
(Fig.~\ref{1body_configs}E-H). In all cases, the global minimum occurs at $\theta=\pi/2$
(Fig.~\ref{1body_configs}J), i.e. when the ellipsoid long axis is oriented perpendicularly to $\dir$, but the
configuration at $\theta=0$ enjoys some metastability, especially as $k$ increases. The metastable ($\theta =0$)
and the stable ($\theta = \pi/2$) solutions are separated by a free energy barrier at $\theta \simeq \pi/12\,$.
For a slight ellipsoid rotation away from $\theta =0$, this free energy barrier may arise from a large
elongation of the defect line and the corresponding increase of the line free energy
(Fig.~\ref{1body_configs}B,C), which, at such small $\theta$ values, is not counterbalanced by any relaxation of
elastic distortions in the bulk. The total free energy of the system is indeed dominated by bulk director
distortions. The height of the energy barrier grows with the ellipsoid aspect ratio $k$ (Fig.~\ref{en_vs_k},
open symbols) and amounts to $\approx 10^3\,k_BT$ for micron-sized particles with $k=7\,$. As $\theta$ increases
further, the above scenario reverses -- the defect line still grows, but at a smaller ``rate'' and the additional
line energy penalty does not prevent the total free energy to decrease steadily, via the relaxation of the bulk
director distortions, all the way down to $\theta=\pi/2\,$.

 Tip rings configurations are more stable than midplane ones at small $\theta$, but their free energy
grows with $\theta\,$. Tilting the particle away from $\theta=0$ indeed generates longer tip rings (see
Fig.~\ref{1body_configs}F-H) and the total free energy rises. Such conformations eventually loose their
stability for $\theta>3\pi/8$ as the corresponding energy branches suddendly fall down and join the ones
computed for midplane rings (Fig.~\ref{en_vs_k}, solid symbols).

Besides midplane and tip rings, a surprising metastable double ring configuration also emerges out of the
computations, but only in a narrow $\theta$ range ($\theta\in[1.04,1.22]\,$rad) (see Fig.~\ref{1body_configs}I).
The associated free energy curve (not shown) is always located above the curves computed for midplane and tip
rings. This double ring conformation has an interesting topological property. One of the rings (the bottom one
in Fig.~\ref{1body_configs}I) has the director structure of a tip ring and therefore carries a topological point
charge equal to $-1\,$. Consequently, the second ring must have zero topological point charge. A fine inspection
of the director topology around the second loop reveals a hybrid structure composed of two parts characterised
by winding numbers equal to $-1/2$ and $+1/2\,$. {Insets in Fig.~\ref{1body_configs}I show the details of
the director configurations around the $+1/2$ (red box) and  $-1/2$ (blue box) disclination profiles. The
transition between the two proceeds by rotating the director by $\pi$ around an axis perpendicular to the
disclination line (see figure 2 in \cite{copar:13}).} Such a peculiar topology indeed yields zero topological
point charge, as explained in \cite{copar:13}. \\

\begin{figure}
\centering
\includegraphics[width=0.8\textwidth]{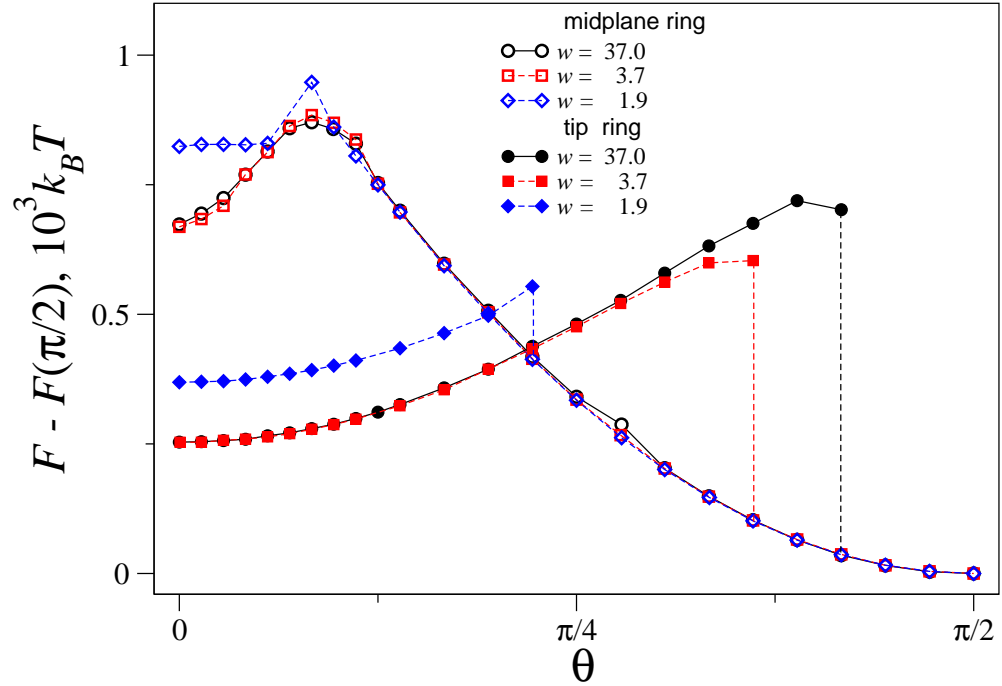}
\caption{Landau-de Gennes free energy as a function  of the angle $\theta$ between the
far-field director and the long axis of the ellipsoid for several values of the dimensionless
homeotropic anchoring strength $w$ and  $k = 5\,$. Open (solid) symbols correspond to
 the free energy branches of midplane ring (tip ring) configurations.} \label{en_vs_W}
\end{figure}

\begin{figure}
\centering
\includegraphics[width=0.8\textwidth]{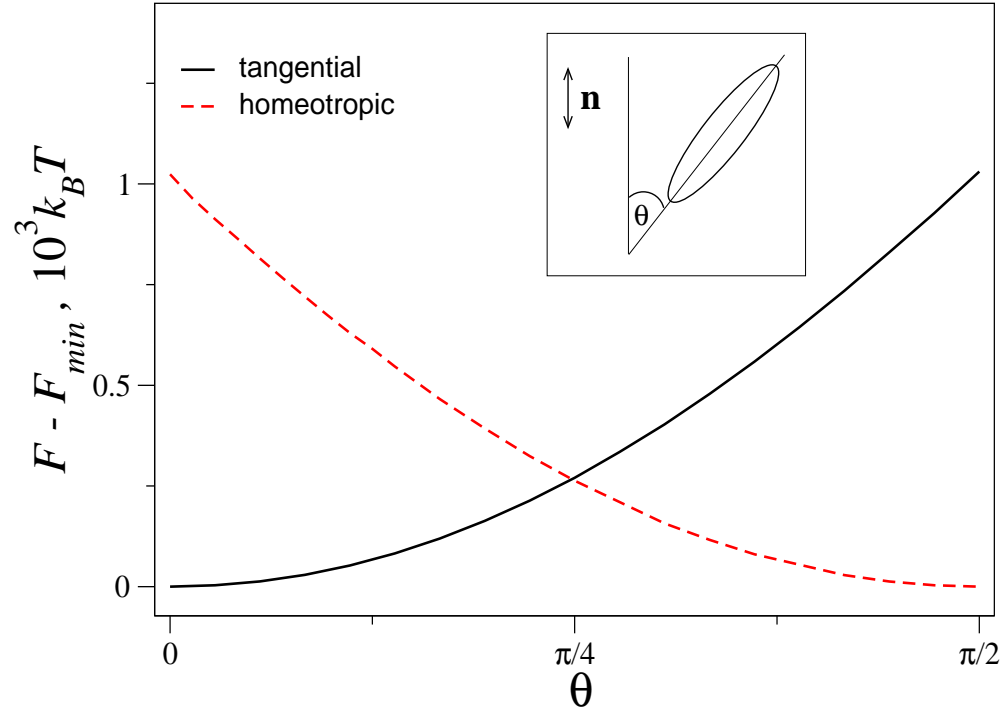}
\caption{Landau-de Gennes free energy as a function of $\theta$  for weak tangential
 (black solid line) and weak homeotropic (red dashed line) anchorings. For strong
tangential anchoring the free energy curve has the same shape as the black solid
curve plotted here (the free energy values are only shifted a bit). Aspect ratio $k = 5$;
dimensionless anchoring strength $w = 0.4$.} \label{weak_anchoring}
\end{figure}

\begin{figure}
\centering
\includegraphics[width=0.9\textwidth]{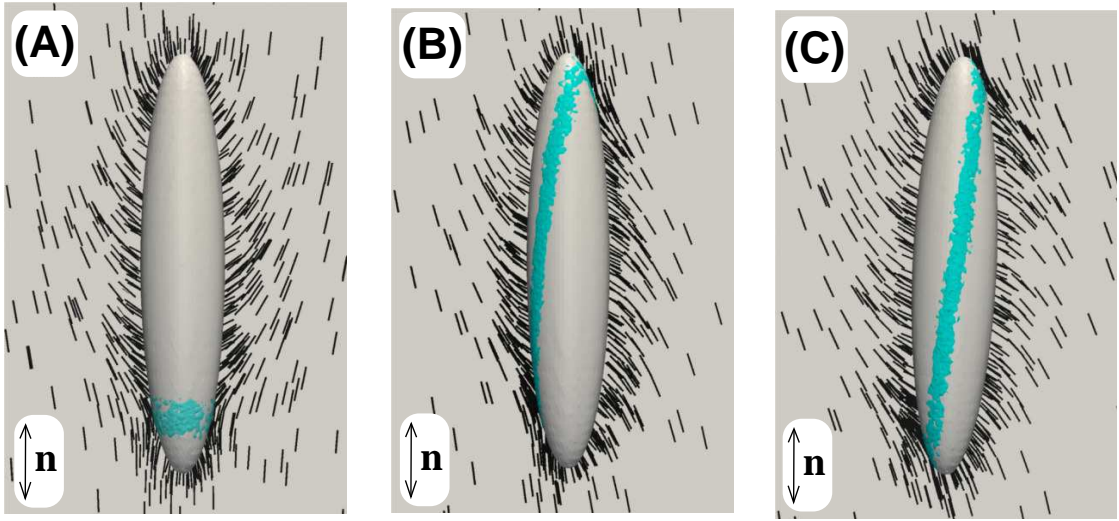}
\caption{Director configurations around an ellipsoidal particle,  $k = 5$, with weak homeotropic anchoring ($w =
0.4$). \textbf{(A)}: remnant of the tip ring configuration, $\theta =0$;  \textbf{(B)}: director configuration
which is not observed for larger $w$ and could be considered as a remnant of the stretched midplane ring
configuration, $\theta = 0^\circ$; \textbf{(C)}: $\theta = 2.5^\circ$. Isosurfaces of constant scalar order
parameter, $Q = 0.97 Q_b\,$, are shown in blue, where $Q_b$ is the bulk value of the scalar order parameter.} \label{1body_conf_weak_homeo}
\end{figure}

\noindent \textit{Weak homeotropic anchoring} -- Decreasing the (dimensionless) strength $w$ of the homeotropic anchoring
alters significantly the free energy landscape of an isolated ellipsoid, as shown in Fig.~\ref{en_vs_W}. For
midplane rings, weaker anchoring tends to suppress the energy barrier which separates the states $\theta=0$ and
$\theta=\pi/2\,$. The angular domain of existence of tip rings shrinks from $\Delta\theta\simeq 1.3\,$rad
($w=37$) down to $\Delta\theta\simeq 0.7\,$rad ($w=1.9$). The surface director deviates more and more from the
local surface normal to minimize the bulk elastic distortions. In this case, the contribution from the surface
free energy $F_s$ increases and, correspondingly, that of the bulk free energy $F$ decreases. For even lower
values of $w$ ($=0.4$), tip rings and double rings no longer exist, the director field becomes nonsingular
and there is only one branch in the free energy which decreases monotonously from $\theta=0$ down to
$\theta=\pi/2$ (see dashed curve in Fig.~\ref{weak_anchoring}). At $\theta=0\,$, two director configurations are
possible (the difference in free energy is of the order of the numerical error), although they are both unstable
with respect to rotations. One of the configurations has a broken up-down symmetry (see
Fig.~\ref{1body_conf_weak_homeo}A) and is a remnant of the tip ring configurations. The second configuration,
shown in Fig.~\ref{1body_conf_weak_homeo}B, has the director structure reminescent of that of stretched midplane
rings (see Fig.~\ref{1body_configs}C), but without disclination lines. Similar director configurations are also
obtained for $\theta>0\,$, (Fig.~\ref{1body_conf_weak_homeo}C). \\

\noindent \textit{Tangential anchoring} -- Switching to tangential anchoring conditions yields a less rich behavior.
Whatever the anchoring strength, be it strong ($w=37$) or weak ($w=0.4$), the free energy curve always decreases
monotonously from $\theta=\pi/2$ down to $\theta=0\,$, which is the only stable state
(Fig.~\ref{weak_anchoring}). Hence, in this case, the ellipsoid will orient its long axis parallel to $\dir$, in
contrast to what is predicted for homeotropic anchoring. Therefore, the nature of the anchoring seems to control
the preferential orientation of the ellipsoids in the nematic phase.

\begin{figure}
\centering
\includegraphics[width=0.9\textwidth]{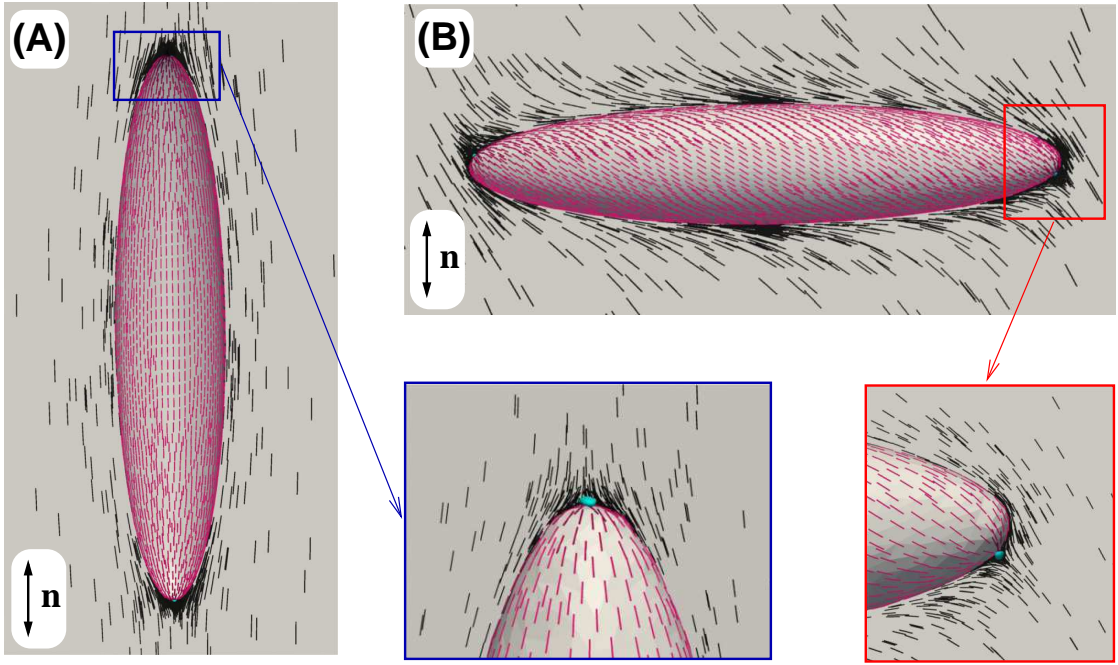}
\caption{ Director configurations, ${\bf n}({\bf r})$, in the bulk (black rods) and $\bf {n}_s(\bf {r})$ on the
surface (magenta rods) of an ellipsoidal colloidal particle with $k = 5$, and strong tangential  anchoring ($w
=37$). \textbf{(A)}:  $\theta =0$;  \textbf{(B)}: $\theta = \pi/2\,$. Insets: zoomed-in views of  ${\bf n}({\bf
r})$ and ${\bf n}_s({\bf r})$ near boojums at the tips of the particles. Isosurfaces of constant scalar order
parameter, $Q =0.6 Q_b\,$, are shown in blue.} \label{1body_conf_planar}
\end{figure}

For strong tangential anchoring, the director field exhibits the typical surface defects known as boojums at the
ellipsoid tips (see snapshot on Fig.~\ref{1body_conf_planar}). This configuration is similar to that computed in
\cite{mondi} using an approximate two-dimensional approach. For weaker anchoring ($w=0.4$), the boojums
disappear (not shown here).

Note that for strong tangential anchoring, a quadratic dependence of the elastic energy $F_{el}$ on $\theta$ for a
rod dispersed in a nematic was predicted \cite{broc} and verified experimentally \cite{lapo04}. From
Fig.~\ref{weak_anchoring}, it seems like such a law, i.e. $F_{el}\propto \theta^2$, holds as well in the weak
anchoring regime (see the Supplementary Information for more details). \\

\subsubsection{Experimental observations}

\noindent In this paragraph, we briefly describe a few qualitative experiments that support the above numerical results.
We mainly focus on the influence of the anchoring type at the LC-colloid interface.

\begin{figure}
\centering
\includegraphics[width=0.9\textwidth]{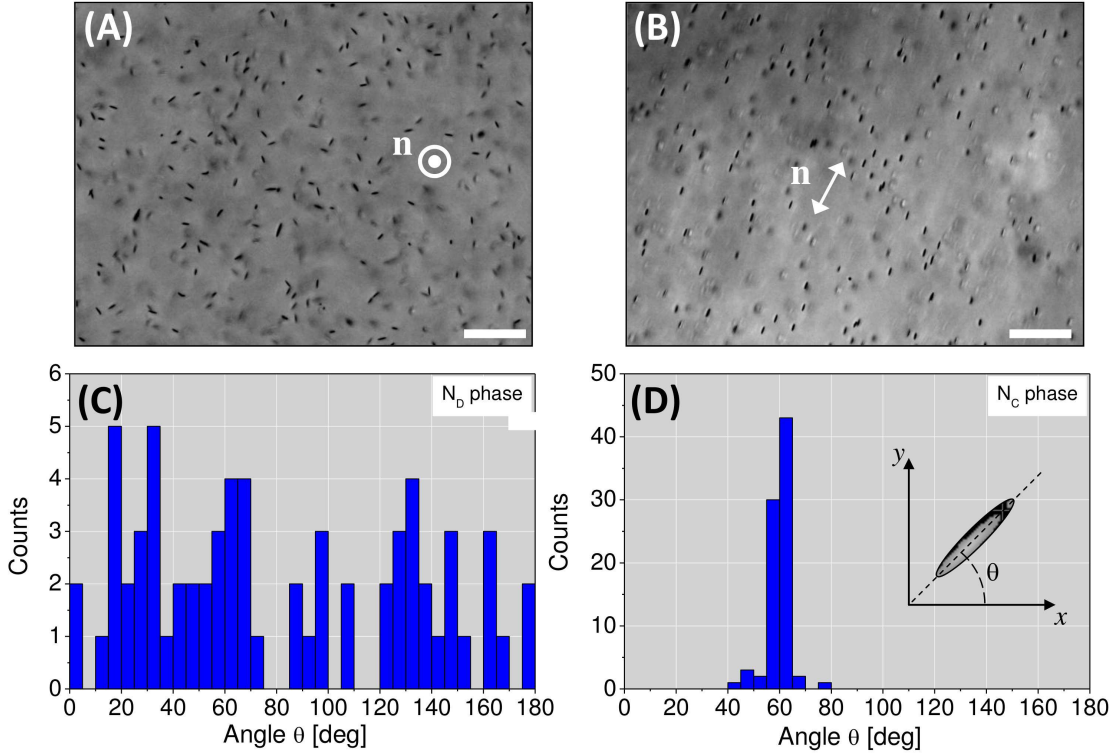}
\caption{\textbf{(A)} and \textbf{(B)} Optical microscopy images of ellipsoids with aspect ratio $k=8.3$
dispersed in the $\ND$ phase \textbf{(A)} and $\NC$ phase \textbf{(B)}. The images were recorded one day after sample preparation
in 1 mm thick cells with a particle mass fraction of 0.01\%. Scale bars: $60\,\mu$m. \textbf{(C)} and
\textbf{(D)} Corresponding histograms of the angular distribution $\theta$ of the ellipsoids in the observation
$xy$-plane for both the $\ND$ \textbf{(C)} and $\NC$ \textbf{(D)} phases. In the latter case, the angular distribution has a
pronounced peak around a well-defined angle and the ellipsoids' long axes are well-oriented parallel to the
local director $\dir\,$. In the $\ND$ phase, the ellipsoids long axes are oriented perpendicular to $\dir$ with
a random orientation in the $xy$-plane since the angular distribution does not exhibit any peak.}
\label{exp_fig1}
\end{figure}

In a previous report \cite{mondi}, we experimentally investigated the behavior of micrometer-sized prolate
ellipsoids, of aspect ratio $k\,$, embedded in a lyotropic $\NC$ phase consisting of an aqueous solution of
rod-like micelles. For a given particle concentration $\phi_P$ ($=0.01\,$wt.\%), it was shown that short-$k$
ellipsoids, just like spheres, aggregate and form anisotropic structures oriented at an angle with respect to
the local background director, which are typical of quadrupolar elastic interactions. However, this is no longer
the case when $k$ reaches a well-defined value $k_C\simeq 4.3$ (which depends on $\phi_P\,$, as discussed
further): above $k_C\,$, the ellipsoids remain homogeneously dispersed and apparently do not interact one
another, even over long periods of time (several months). This is recalled in Fig.~\ref{exp_fig1}B where the non
aggregated ellipsoids ($k=8.3$) are aligned along the local director field, with a tangential anchoring of the
rod-like micelles at the particles' surfaces. The histogram of the angular distribution $\theta$ (see scheme in
Fig.~\ref{exp_fig1}D) has indeed a well-marked peak along the nematic director $\dir$ \cite{contpg}. However,
this is not so when the same ellipsoids are embedded in the $\ND$ phase of the lyotropic system where the
anchoring of the disk-like micelles is now homeotropic \cite{poul99}. On an average, the long axes of ellipsoids
lie in planes oriented perpendicularly to $\dir\,$. Within each plane, the orientation of each ellipsoid is
random as is evidenced by the broad angular distribution histogram of Fig.~\ref{exp_fig1}C.

These observations agree qualitatively well with the calculations described above: the experiments confirm that
the nature of the anchoring controls the ellipsoids' average orientation with respect to the far-field director
(see again Fig.~\ref{exp_fig1} and Fig.~\ref{weak_anchoring}).

Furthermore, in the $\ND$ phase, almost all ellipsoids have their long axes normal to $\dir$
(Fig.~\ref{exp_fig1}A). If the anchoring were strong, a significant amount of particles should have been
observed with their long axes parallel to $\dir$ or slightly tilted, according to the metastable states
predicted by the computations (Figs.~\ref{en_vs_k} and \ref{en_vs_W}). No such metastable states exist in the
graph of Fig.~\ref{weak_anchoring} obtained in the weak anchoring regime, which is then more in line with the
experimental situation. Hence, these results suggest that the anchoring strength is probably weak in our
experiments. With $w$ given by $w=WQ_b^2R_0/K_2$ (Supplementary Information) and taking the following values
$w\sim 1\,$, $K_2\simeq K\sim 1\,$pN, $R_0\sim\mu$m, $Q_b\sim 0.4\,$, one can get an estimate of the anchoring
strength $W\simeq 10^{-6}\,\mathrm{J.m}^{-2}\,$, which is a reasonable value for a weak anchoring case
\cite{lock,pier,lavr98}. \\

\subsection{Pairwise interaction and collective behavior}

\noindent We here provide a few additional experimental and computational results pertaining to a collection of
ellipsoids.

\begin{figure}
\centering
\includegraphics[width=0.9\textwidth]{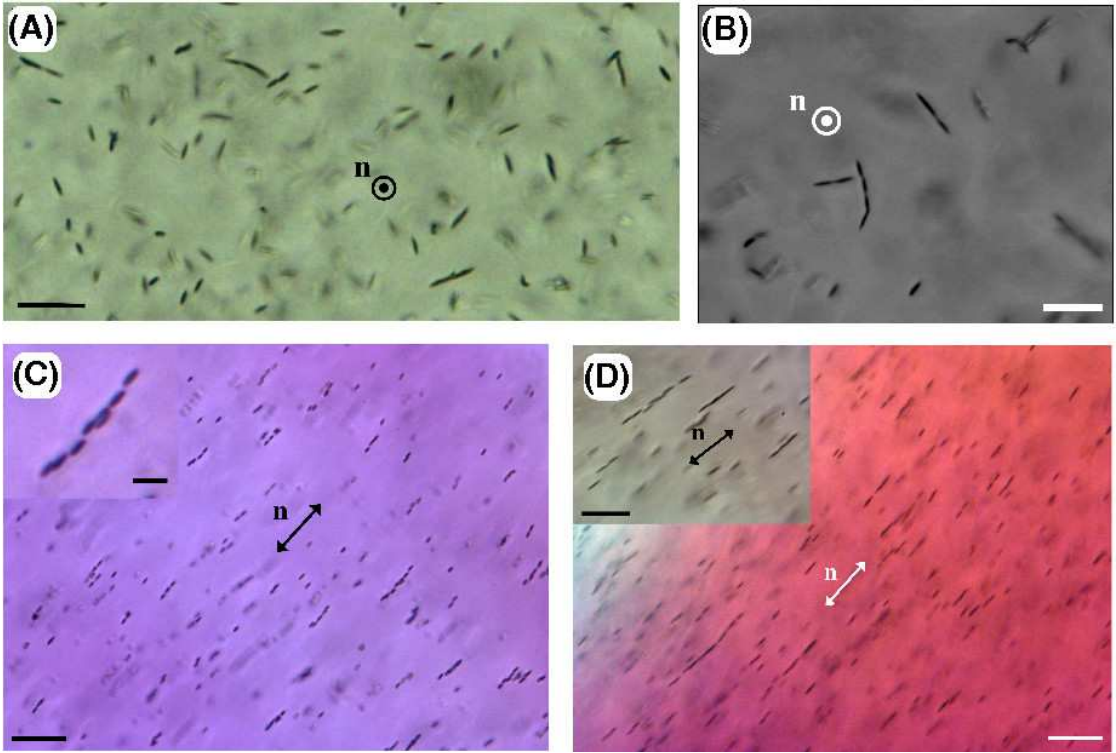}
\caption{Optical microscopy images of ellipsoids dispersed in nematic phases. \textbf{(A)} and \textbf{(B)} $\ND$
phase with aspect ratio $k=8.3\,$. The particles have their long axis oriented perpendicular to the director
$\dir$ and preferentially aggregate tip-to-tip. \textbf{(A)} Scale bar: $19.5\,\mu$m. \textbf{(B)} Scale bar:
$25\,\mu$m. The images were recorded 10 days after sample preparation in 1 mm thick cells with a particle mass
fraction of 0.01\%. \textbf{(C)} and \textbf{(D)} $\NC$ phase with aspect ratios $k=4.3$ \textbf{(C)} and $k=8.3$
\textbf{(D)}. The ellipsoids tend to form (short) chains in which the particles are mainly aggregated in a
tip-to-tip manner. The chains align along the local director $\dir$ (double arrows). \textbf{(C)} Aspect ratio
$k=4.3\,$. Scale bar: $25\,\mu$m. Inset: zoomed-in view of a linear aggregate. Scale bar: $16\,\mu$m.
\textbf{(D)} Aspect ratio $k=8.3\,$. Scale bar: $29\,\mu$m. Inset: close-up view of linear aggregates. Scale
bar: $8.5\,\mu$m. The images were recorded one month after sample preparation in 1 mm thick cells with a
particle mass fraction of 0.04\%. } \label{exp_fig2}
\end{figure}

\begin{figure}
\centering
\includegraphics[width=0.8\textwidth]{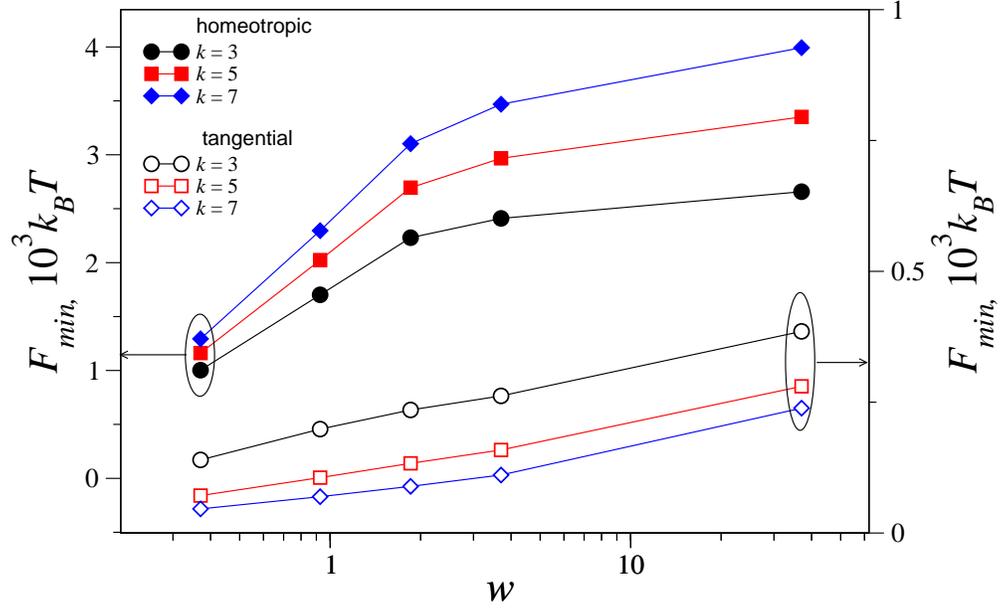}
\caption{Variation of the free energy minimum of a single ellipsoid as a function of the anchoring strength for
several aspect ratios and two different anchoring conditions. For homeotropic anchoring, $\theta=\pi/2\,$, while
for tangential anchoring, $\theta=0\,$.} \label{1body_en_min_vs_W-k}
\end{figure}

In the $\ND$ phase, our observations reveal that, whatever the aspect ratio $k\in]1,10]\,$, the ellipsoids
cluster within comparable timescales than the starting microspheres at the same particle concentration
($0.01\,$wt.\%). This is shown in Fig.~\ref{exp_fig2}A,B where short chains of ellipsoids are oriented
perpendicularly to $\dir\,$, and within which the particles contact in a tip-to-tip manner. Consequently, it
turns out that, in the $\ND$ phase, it is not possible to define a critical aspect ratio above which the
ellipsoids can remain homogeneously dispersed, at least within the probed $k$ range. A possible explanation for
this may be found by examining how the total free energy (Eq.~(\ref{free_energy})) of a single ellipsoid varies
as a function of $k$ with homeotropic anchoring conditions ($\theta=\pi/2$). Such a graph is plotted in
Fig.~\ref{1body_en_min_vs_W-k} (solid symbols). Whatever the anchoring strength, the longer the ellipsoid, the
higher the free energy cost. The length of the Saturn ring defect also increases with $k$, but the defect core
free energy always remains much smaller than the elastic free energy due to out-of-core director field
distortions (results not shown). Hence, aggregation is expected for any $k$ and, as mentioned above, this is
indeed observed experimentally.

This behavior differs drastically from that evidenced in the $\NC$ phase where the anchoring is tangential.
Indeed, the graph on Fig.~\ref{1body_en_min_vs_W-k} (open symbols) now exhibits the opposite trend: as $k$
increases, the value of the free energy minimum decreases. The director topology only features boojums defects
and no disclination loop (see Fig.~\ref{1body_conf_planar}). These computations strongly support previous
calculations in 2D where the same trend was evidenced (see Fig.~3 in \cite{mondi}). Long-$k$ ellipsoids cost
less free energy than short ones and therefore become easier to disperse in the nematic matrix because of weaker
elastic distortions.

Overall, these results highlight again the importance of the anchoring type at the particle-nematic interface.

It is worth mentioning that structures similar to those exhibited in Fig.~\ref{exp_fig2}A,B were experimentally
reported with cylinders \cite{tkal08} and theoretically predicted for spherocylinders \cite{hung} in nematics
with homeotropic anchoring conditions. In such studies, the director topology around the inclusions features a
Saturn ring defect consisting in a disclination loop surrounding the particle long axis, which is normal to
$\dir\,$. Using the same anchoring type, our calculations also predict a Saturn ring defect which we have called
midplane ring (Fig.~\ref{1body_configs}A-D). But, as aforesaid, other peculiar director morphologies are
possible (Fig.~\ref{1body_configs}E-I).

\begin{figure}
\centering
\includegraphics[width=0.8\textwidth]{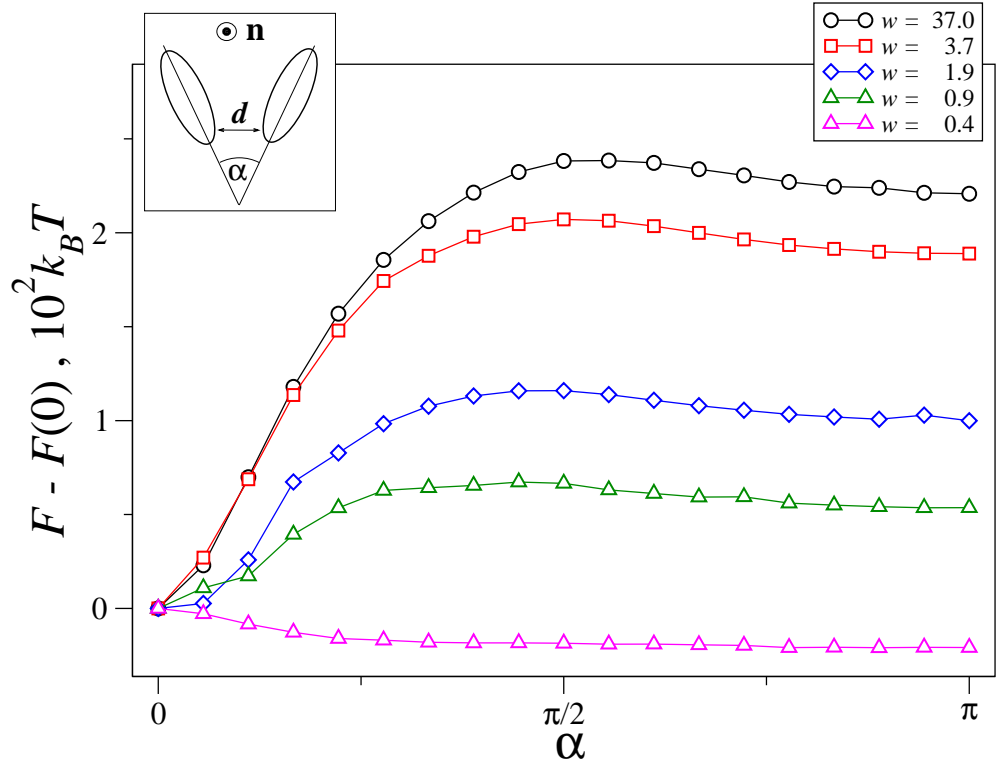}
\caption{Effective pair interaction as a function of the relative colloidal orientation $\alpha$, at fixed
surface-to-surface distance $d =0.1 B$ for several values of $w$ (homeotropic anchoring). $B=10\xi\approx
10\,\mu$m is the azimuthal (short) radius of the ellipsoids' surfaces. The ellipsoids' long axes are confined in
a plane perpendicular to the far-field director $\dir$. Aspect ratio $k = 5\,$.} \label{2body_en_vs_alpha}
\end{figure}

\begin{figure}
\centering
\includegraphics[width=0.9\textwidth]{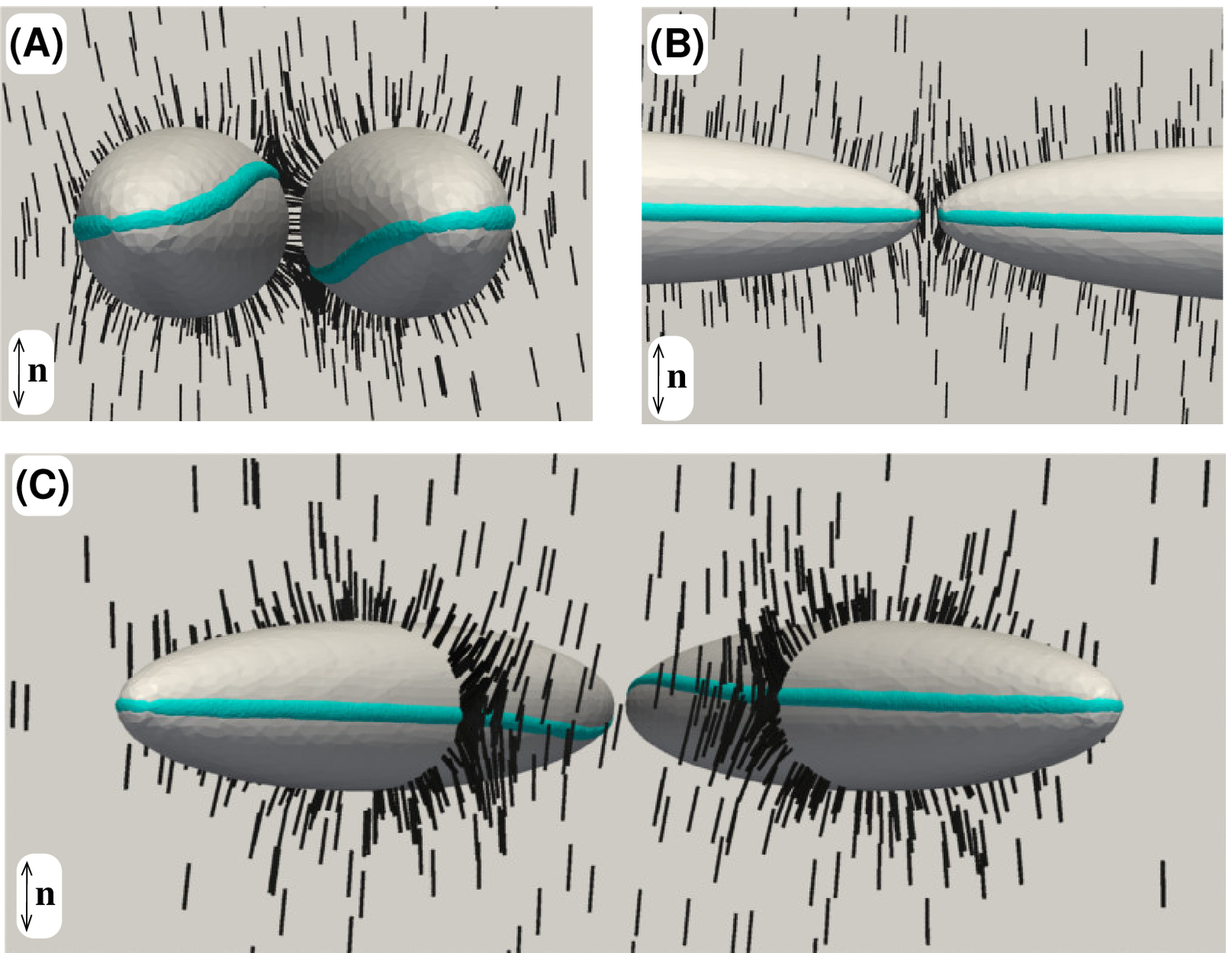}
\caption{{Director configurations (black rods) around two ellipsoidal particles,  $k = 5$, with weak homeotropic anchoring ($w = 1.9$). \textbf{(A)}:  $\alpha =0$;  \textbf{(B)}: $\alpha =180^\circ$; \textbf{(C)}: $\alpha =50^\circ$.
Isosurfaces of constant scalar order parameter, $Q = 0.6 Q_b\,$, are shown in blue, where $Q_b$ is the bulk
value of the scalar order parameter.}} \label{2body_conf_weak_homeo}
\end{figure}

To gain further insights on the anisotropic structures reported on Fig.~\ref{exp_fig2}A,B, we have computed the
effective pair interaction as a function of the relative colloidal orientation defined by angle $\alpha$ (see
Fig.~\ref{2body_en_vs_alpha} and inset). The calculations were carried out for a fixed surface-to-surface
distance $d$ and several values of the homeotropic anchoring strength $w\,$. It is found that strong anchoring
favors the side-to-side configuration with $\alpha=0\,$. But the tip-to-tip configuration ($\alpha=\pi$) seems
to enjoy some metastability since the free energy curve is non monotonous and features a maximum around
$\pi/2\,$. As the anchoring strength decreases, so does the free energy difference between the side-to-side and
the tip-to-tip configurations. For the lowest considered value of $w$ ($w=0.4$), the free energy minimum is
reached for the tip-to-tip geometry although the free energy values do not differ greatly for
$\alpha\in[\pi/2,\pi]\,$. {The large $\alpha$ repulsive ``tails'' of the interaction profiles for strong to
moderate anchoring (Fig.~\ref{2body_en_vs_alpha}) can be understood through an electrostatic analogy as the
repulsion of two elastic quadrupoles \cite{poul97,lube}. The emergence of the free energy barriers, and attractive forces at
smaller values of $\alpha\,$, is due to the so-called defects sharing mechanism first described in
\cite{tasinkevych:2002}. At small angular separations, the disclination lines in the inner regions start to
deform in the opposite directions (see Fig.~\ref{2body_conf_weak_homeo}C), thereby minimizing the overall free
energy and leading to the elastically bonded configuration at $\alpha =0$ (see
Fig.~\ref{2body_conf_weak_homeo}A). For weak anchoring, the free energy is dominated by the surface anchoring
term, which is minimized for the tip-to-tip orientation (Fig.~\ref{2body_conf_weak_homeo}B).} Experimentally,
only tip-to-tip structures are observed, and therefore, these calculations again suggest that the anchoring of
the micelles on the ellipsoids surfaces is probably weak. The same conclusion was indeed already inferred above
when comparing the observations of Fig.~\ref{exp_fig1} with the computations reported on Figs.~\ref{en_vs_W} and
\ref{weak_anchoring}.  {Finally, we note that the computed effective elastic forces acting upon ellipsoids
in the side-to-side configuration ($\alpha=0$, strong anchoring) are of the order of $1\,$pN, which correlates
well with earlier experiments on micrometer-sized ``homeotropic'' spheres for which radial elastic forces of
about $\sim 10\,$pN were reported \cite{smalyukh:2005}.}

Lastly, we have briefly investigated the effect of a four-fold increase of the particle concentration in the
$\NC$ phase, i.e. with $\phi_P=0.04$ wt.\%. Ellipsoids with aspect ratios $k=4.3,\,5.8$ and 8.3 remained
well-dispersed for about three weeks. But one month after sample preparation, anisotropic chains parallel to
$\dir\,$ could be clearly distinguished as seen in Fig.~\ref{exp_fig2}C,D. Inside the chains, the ellipsoids
preferentially contact by their tips, like in the aggregates that form in the $\ND$ phase
(Fig.~\ref{exp_fig2}A,B). Hence, increasing $\phi_P$ leads to particle aggregation which is not too surprising
since the mean distance between inclusions decreases, thereby making them more sensitive to attractive elastic
interactions. However, the strength of these interactions was predicted to be a decreasing function of $k$ for
pairs of ellipsoids \cite{mondi}. One may therefore surmise that the critical aspect ratio $k_C$ varies with
$\phi_P$, and furthermore, that it is an increasing function of $\phi_P$. However, a great deal of additional
experiments would be required to test this conjecture. \\

\section{Conclusion}

\noindent To conclude, we have studied through both numerics and experiments the static behavior of elongated particles
embedded in a nematic phase. Among the most salient results of our investigation, let us point out that (i) the
nature of the anchoring at the LC-colloid interface controls the preferential orientation of ellipsoids with
respect to the far field nematic director, (ii) the strength of the anchoring has a direct impact on the packing
geometry of pairs of ellipsoids, (iii) several metastable states exist in the free energy landscape when tilting
the ellipsoid long axis with respect to $\dir\,$; some of these states exhibit unusual topological properties
and director fields; and (iv) no ``critical aspect ratio'' could be evidenced in the $\ND$ phase: the ellipsoids
are always found to aggregate whatever their aspect ratio, in fierce contrast to what was previously observed in
the $\NC$ phase.

Our investigation therefore brings out additional insights about the influence of particle geometry on the
behavior of LC colloids. Optimising the shape of colloidal particles together with their surface properties is certainly a
worthy route to be explored further in view of achieving useful homogeneous LC dispersions with high particle
loads. On a more fundamental basis, composites such as LC colloids also provide an ideal playground to
deliberately nucleate various kinds of defects and study their interactions, which is of great interest from a
theoretical perspective \cite{alex}. \\

\section{Methods}

\noindent {\bf Preparation of particles}. Prolate ellipsoidal particles were prepared using the process reviewed in
\cite{cham}. Briefly, the latter consists in a uniaxial mechanical stretching of polymeric spherical particles
at constant volume, which are initially trapped in a film-forming polymer matrix. In our experiments, the matrix
was made of 89\%-hydrolysed polyvinyl alcohol (PVA -- $120,000\leq M_W\leq 160,000\,\mbox{g.mol}^{-1}$ from
Fluka\copyright) whereas the starting particles consisted of $2\,\mu$m-diameter polystyrene (PS) beads
(Polysciences\copyright). Monodisperse prolate ellipsoids of controllable aspect ratio $k=A/B\,$ ranging from 1
to 10 are easily achievable. We assume that the ellipsoids are cylindrically symmetric and defined by semi-axes
$(A,B,B)\,$, where $A$ (resp. $B$) is the ellipsoid semi-long (resp. semi-short) axis. The resulting particles
are water-dispersable and aqueous suspensions may be stored for several months without significant aggregation.
However, despite numerous washing cycles, some PVA inevitably remains adsorbed at the particles surface.
Although we did not quantify this adsorption, we checked that PVA-coated PS spheres -- i.e. spheres that
underwent the same preparation procedure as ellipsoids, but without the stretching step, -- behaved like the
bare PS beads when dispersed in nematic phases as described below. Therefore, a PVA-coating had no effect on the
particles behavior, certainly because of the entropic nature of the anchoring in lyotropic nematics (see
hereafter). \\

\noindent{\bf Liquid crystal}. We used a lyotropic liquid crystal system composed of sodium dodecyl sulfate (SDS),
decanol (DeOH) and water as the dispersing solvent for the ellipsoids. The phase diagram of this mixture
exhibits two nematic domains \cite{quist,nesr}: a calamitic nematic phase  ($\NC$), made of 9 nm-long and 3.6
nm-wide rod-like surfactant micelles; and a discotic nematic phase ($\ND$), composed of 8 nm-diameter and 3.5
nm-thick disk-like micelles. We chose the following two compositions: one in the $\NC$ phase (water, 71\%; SDS,
24.5\%; DeOH, 4.5\% by weight) and the other in the $\ND$ phase (water, 73\%; SDS, 23.5\%; DeOH, 3.5\% by
weight). The $\NC$ phase was stable from $14^\circ$C, the crystallisation temperature of SDS, up to $32^\circ$C
where it transited to an isotropic micellar phase ($\mathrm{L}_1$). The $\ND$ phase transited to the $\NC$ phase
below $20^\circ$C, and was stable over a large temperature range. Because of the high content of SDS, which is
at least 140 times the critical micellar concentration (CMC) of SDS
($\mathrm{CMC}_{\mathrm{SDS}}=8.10^{-3}\,$M), the ionic strength of the nematic solutions is very high. As a
consequence, the electrostatic interactions are strongly screened. Indeed, with the above concentrations, the
debye screening length, $\kappa^{-1}\,$, is of order $\sim 0.1\,$nm \cite{pash,monda96}. This estimate remains
valid for the two nematic domains of the ternary mixture since SDS concentrations range from 140 CMC up to about
200 CMC. Hence, the electrostatic repulsions are only effective at very short range, resulting in hard-core type
interactions between micelles and between micelles and walls. Orientation of anisotropic micelles close to walls
is then mainly driven by entropy: as shown previously, the micelles will tend to align parallel to surfaces to
minimize their excluded volume \cite{poni}. This preferential alignment leads to tangential (resp. homeotropic)
anchoring of the micelles on surfaces in the $\NC$ (resp. $\ND$) phase. Tuning of the anchoring conditions may
then be achieved by introducing the particles either in the $\NC$ phase (tangential case) or in the $\ND$ phase
(homeotropic case) \cite{poul99}.\\

\noindent {\bf Samples}.
 Ellipsoids were suspended in nematic phases at mass fractions ranging from 0.01\% to 0.06\%. At such low
concentrations, the lyotropic nematic phases were not altered by the presence of the particles. Unlike
thermotropic nematic dispersions, no further chemical functionalization of the particles surface was required
here since the particles are water-dispersable. The nematic suspensions were put in 1 mm-thick optical glass
cells which were further capped and sealed to prevent evaporation. The samples were thermostated at $24^\circ$C,
and observed using standard polarizing optical microscopy.\\

\noindent{\bf Acknowledgements}

\noindent We acknowledge financial support from the French government, the Conseil R\'{e}gional d'Aquitaine and
the Agence Nationale de la Recherche under grant No. PACTIS JC07-198199. M. T. acknowledges the 7th Framework
International Program Research Staff Exchange Scheme Marie-Curie Grant PIRSES-GA-2010-269181. Part of
this work was also done in the frame of the ITN-COMPLOIDS European network.\\

\newpage

\begin{thebibliography}{99}

%\expandafter\ifx\csname natexlab\endcsname\relax\def\natexlab#1{#1}\fi \expandafter\ifx\csname
%bibnamefont\endcsname\relax
%  \def\bibnamefont#1{#1}\fi
%\expandafter\ifx\csname bibfnamefont\endcsname\relax
%  \def\bibfnamefont#1{#1}\fi
%\expandafter\ifx\csname citenamefont\endcsname\relax
%  \def\citenamefont#1{#1}\fi
%\expandafter\ifx\csname url\endcsname\relax
%  \def\url#1{\texttt{#1}}\fi
%\expandafter\ifx\csname urlprefix\endcsname\relax\def\urlprefix{URL }\fi \providecommand{\bibinfo}[2]{#2}
%\providecommand{\eprint}[2][]{\url{#2}}

   \bibitem{bury} Burylov S. and Raikher Y. L. Orientation of a solid particle embedded in a monodomain nematic liquid crystal. \textit{Phys. Rev. E} \textbf{50}, 358-367 (1994).

\bibitem{tere} Terentjev E. M. Disclination loops, standing alone and around solid particles, in nematic liquid crystals. \textit{Phys. Rev. E} \textbf{51}, 1330-1337 (1995).

   \bibitem{ruhw} Ruhwandl R. W. and Terentjev E. M. Monte Carlo simulation of topological defects in the nematic liquid crystal matrix around a spherical colloid particle. \textit{Phys. Rev. E} \textbf{56}, 5561-5565 (1997).

   \bibitem{poul97} Poulin P., Stark H., Lubensky T. C. and Weitz D. A. Novel Colloidal Interactions in Anisotropic Fluids. \textit{Science} \textbf{275}, 1770-1773 (1997).

   \bibitem{poul98} Poulin P. and Weitz D. A. Inverted and multiple nematic emulsions. \textit{Phys. Rev. E} \textbf{57}, 626-637 (1998).

   \bibitem{lube} Lubensky T. C., Pettey D., Currier N. and Stark H. Topological defects and interactions in nematic emulsions. \textit{Phys. Rev. E} \textbf{57}, 610-625 (1998).

   \bibitem{monda99} Mondain-Monval O., Dedieu J. C., Gulik-Krzywicki T. and Poulin P. Weak surface energy in nematic dispersions: Saturn ring defects and quadrupolar interactions. \textit{Eur. Phys. J. B} \textbf{12}, 167-170 (1999).

   \bibitem{star99} Stark H. Director field configurations around a spherical particle in a nematic liquid crystal. \textit{Eur. Phys. J. B} \textbf{10}, 311-321 (1999).

   \bibitem{star01} Stark H. Physics of colloidal dispersions in nematic liquid crystals. \textit{Phys. Rep.} \textbf{351}, 387-474 (2001).

   \bibitem{muse} Mu\v{s}evi\v{c} I., \v{S}karabot M., Tkalec U., Ravnik M. and \v{Z}umer S. Two-Dimensional Nematic Colloidal Crystals Self-Assembled by Topological Defects. \textit{Science} \textbf{313}, 954-958 (2006).

   \bibitem{gu} Gu Y. and Abbott N. L. Observation of Saturn-Ring Defects around Solid Microspheres in Nematic Liquid Crystals. \textit{Phys. Rev. Lett.} \textbf{85}, 4719-4722 (2000).

   \bibitem{grol} Grollau S., Abbott N. L. and de Pablo J. J. Spherical particle immersed in a nematic liquid crystal: Effects of confinement on the director field configurations. \textit{Phys. Rev. E} \textbf{67}, 011702 (2003).

   \bibitem{skar_a} \v{S}karabot M., Ravnik M., \v{Z}umer S., Tkalec U., Poberaj I., Babi\v{c} D., Osterman N. and Mu\v{s}evi\v{c} I. Two-dimensional dipolar nematic colloidal crystals. \textit{Phys. Rev. E} \textbf{76}, 051406 (2007).

   \bibitem{skar_b} \v{S}karabot M., Ravnik M., \v{Z}umer S., Tkalec U., Poberaj I., Babi\v{c} D., Osterman N. and Mu\v{s}evi\v{c} I. Interactions of quadrupolar nematic colloids. \textit{Phys. Rev. E} \textbf{77}, 031705 (2008).

   \bibitem{loud01} Loudet J.-C. and Poulin P. Application of an Electric Field to Colloidal Particles Suspended in a Liquid-Crystal Solvent. \textit{Phys. Rev. Lett.} \textbf{87}, 165503 (2001).

   \bibitem{fuku} Fukuda J. and Yokoyama H. Stability of the director profile of a nematic liquid crystal around a spherical particle under an external field. \textit{Eur. Phys. J. E} \textbf{21}, 341-347 (2006).

   \bibitem{rama} Ramaswamy S., Nityananda R., Raghunathan V. A. and Prost J. Power-law forces between particles in a nematic. \textit{Mol. Cryst. Liq. Cryst.} \textbf{288}, 175-189 (1996).

   \bibitem{poul97b} Poulin P., Cabuil V. and Weitz D. A.  Direct Measurement of Colloidal Forces in an Anisotropic Solvent. \textit{Phys. Rev. Lett.} \textbf{79}, 4862-4865 (1997).

   \bibitem{smal05} Smalyukh I. I., Lavrentovich O. D., Kuzmin A. N., Kachynski A. V. and Prasad P. N. Elasticity-Mediated Self-Organization and Colloidal Interactions of Solid Spheres with Tangential Anchoring in a Nematic Liquid Crystal. \textit{Phys. Rev. Lett.} \textbf{95}, 157801 (2005).

   \bibitem{taka} Takahashi K., Ichikawa M. and Kimura Y. Force between colloidal particles in a nematic liquid crystal studied by optical tweezers. \textit{Phys. Rev. E} \textbf{77}, 020703(R) (2008).

   \bibitem{eska} Eskandari Z., Silvestre N. M., Tasinkevych M. and Telo da Gama M. M. Interactions of distinct quadrupolar nematic colloids. \textit{Soft Matter} \textbf{8}, 10100-10106 (2012).

   \bibitem{loud00} Loudet J.-C., Barois P. and Poulin P. Colloidal ordering from phase separation in a liquid- crystalline continuous phase. \textit{Nature} \textbf{407}, 611 (2000).

   \bibitem{ravn} Ravnik M., \v{S}karabot M., \v{Z}umer S., Tkalec U., Poberaj I., Babi\v{c} D., Osterman N.
   and Mu\v{s}evi\v{c} I. Entangled Nematic Colloidal Dimers and Wires. \textit{Phys. Rev. Lett.} \textbf{99}, 247801 (2007).

   \bibitem{ogny08} Ognysta U., Nych A., Nazarenko V., Mu\v{s}evi\v{c} I., \v{S}karabot M., Ravnik M., \v{Z}umer
   S., Poberaj I. and Babi\v{c} D. 2D Interactions and Binary Crystals of Dipolar and Quadrupolar Nematic Colloids. \textit{Phys. Rev. Lett.} \textbf{100}, 217803 (2008).

   \bibitem{ogny09} Ognysta U., Nych A., Nazarenko V., \v{S}karabot M. and Mu\v{s}evi\v{c} I. Design of 2D Binary Colloidal Crystals in a Nematic Liquid Crystal. \textit{Langmuir} \textbf{25}, 12092-12100 (2009).

   \bibitem{nych} Nych A., Ognysta U., \v{S}karabot M., Ravnik M., \v{Z}umer S. and Mu\v{s}evi\v{c} I. Assembly and control of 3D nematic dipolar colloidal crystals. \textit{Nat. Commun.} \textbf{4}, 1489 (2013).

   \bibitem{tkal11} Tkalec U., Ravnik M., \v{C}opar S., \v{Z}umer S. and Mu\v{s}evi\v{c} I. Reconfigurable Knots and Links in Chiral Nematic Colloids. \textit{Science} \textbf{333}, 62-65 (2011).

   \bibitem{senyuk:13} Senyuk B., Liu Q., He S., Kamien R. D., Kusner R. B., Lubensky T. C. and Smalyukh I. I. Topological colloids. \textit{Nature} \textbf{493}, 200-205 (2013).

   \bibitem{liu:13} Liu, Q. Senyuk, B., Tasinkevych, M. and Smalyukh I. I.  Nematic liquid crystal boojums with handles on colloidal handlebodies. \textit{Proc. Natl. Acad. Sci. USA} \textbf{110}, 9231-9236 (2013).

   \bibitem{lavr} Lavrentovich O. D., Lazo I. and Pishnyak O. P. Nonlinear electrophoresis of dielectric and metal spheres in a nematic liquid crystal. \textit{Nature} \textbf{467}, 947-950 (2010).

   \bibitem{pish} Pishnyak O. P., Shiyanovskii S. V. and Lavrentovich O. D. Inelastic Collisions and Anisotropic Aggregation of Particles in a Nematic Collider Driven by Backflow. \textit{Phys. Rev. Lett.} \textbf{106}, 047801 (2011).

  \bibitem{lint} Lintuvuori J. S., Stratford K., Cates M. E. and Marenduzzo D. Colloids in Cholesterics: Size-Dependent Defects and Non-Stokesian Microrheology. \textit{Phys. Rev. Lett.} \textbf{105}, 178302 (2010).

   \bibitem{cham} Champion J. A., Katare Y. K. and Mitragotri S. Making polymeric micro- and nanoparticles of complex shapes. \textit{Proc. Natl. Acad. Sci. USA} \textbf{104}, 11901-11904 (2003).

   \bibitem{saca10} Sacanna S., Irvine W. T. M., Chaikin P. M. and Pine D. J.  Lock and key colloids. \textit{Nature} \textbf{464}, 575-578 (2010).

   \bibitem{saca11} Sacanna S. and Pine D. J.  Shape-anisotropic colloids: Building blocks for complex assemblies. \textit{Curr. Opin. Colloid Interface Sci.} \textbf{16}, 96-105 (2011).

   \bibitem{liu11} Liu Y. and Zhang X. Metamaterials: a new frontier of science and technology. \textit{Chem. Soc. Rev.} \textbf{40}, 2494-2507 (2011).

   \bibitem{wood} Wood T. A., Lintuvuori J. S., Schofield A. B., Marenduzzo D. and Poon W. C. K. A Self-Quenched Defect Glass in a Colloid-Nematic Liquid Crystal Composite. \textit{Science} \textbf{334}, 79-83 (2011).

   \bibitem{lock} Lockwood N. A., Gupta J. K. and Abbott N. L.  Self-assembly of amphiphiles, polymers and proteins at interfaces between thermotropic liquid crystals and aqueous phases. \textit{Surf. Sci. Rep.} \textbf{63}, 255-293 (2008).

   \bibitem{prat} Prathap Chandran S., Mondiot F., Mondain-Monval O. and Loudet J.-C.  Photonic Control of Surface Anchoring on Solid Colloids Dispersed in Liquid Crystals. \textit{Langmuir} \textbf{27}, 15185-15198 (2011).

   \bibitem{volt} V\"{o}ltz C., Maeda Y., Tabe Y. and Yokoyama H. Director-Configurational Transitions around Microbubbles of Hydrostatically Regulated Size in Liquid Crystals. \textit{Phys. Rev. Lett.} \textbf{97}, 227801 (2006).

\bibitem{koen09a} Koenig Jr. G. M., Ong. R., Cortes A. D., Moreno-Razo J. A., de Pablo J. J. and Abbott N. L. Single nanoparticle tracking reveals influence of chemical functionality of nanoparticles on local ordering of liquid crystals and nanoparticle diffusion coefficients. \textit{Nano Lett.} \textbf{9}, 2794-2801 (2009).

   \bibitem{toma} Tomar V., Roberts T. F., Abbott N. L., Hern\'{a}ndez-Ortiz J. P. and de Pablo J. J. Liquid Crystal Mediated Interactions Between Nanoparticles in a Nematic Phase. \textit{Langmuir} \textbf{28}, 6124-6131 (2012).

   \bibitem{pgg} de Gennes P. G. and Prost J., \textit{The Physics of Liquid Crystals}, Clarendon Oxford 2nd Ed., 1993.
   \bibitem{pier} Pieranski P. and Oswald P., \textit{Liquid Crystals}, GB Science Publishers, Paris, 2002.

   \bibitem{koen09b} Koenig Jr. G. M., de Pablo J. J. and Abbott N. L.  Characterization of the Reversible Interaction of Pairs of Nanoparticles Dispersed in Nematic Liquid Crystals. \textit{Langmuir} \textbf{25}, 13318-13321 (2009).

   \bibitem{mondi} Mondiot F., Prathap Chandran S., Mondain-Monval O. and Loudet J.-C. Shape-Induced Dispersion of Colloids in Anisotropic Fluids. \textit{Phys. Rev. Lett.} \textbf{103}, 238303 (2009).

   \bibitem{smal08} Smalyukh I. I., Butler J., Shrout J. D., Parsek M. R. and Wong G. C. L. Elasticity-mediated nematiclike bacterial organization in model extracellular DNA matrix. \textit{Phys. Rev. E} \textbf{78}, 030701(R) (2008).

   \bibitem{liu10} Liu Q., Cui Y., Gardner D., Li X., He S. and Smalyukh I. I. Self-Alignment of Plasmonic Gold Nanorods in Reconfigurable Anisotropic Fluids for Tunable Bulk Metamaterial Applications. \textit{Nano Lett.} \textbf{10}, 1347-1353 (2010).

  \bibitem{tkal08} Tkalec U., \v{S}karabot M. and Mu\v{s}evi\v{c} I.  Interactions of micro-rods in a thin layer of a nematic liquid crystal. \textit{Soft Matter} \textbf{4}, 2402-2409 (2008).

   \bibitem{lapo09} Lapointe C. P., Mason T. G. and Smalyukh I. I.  Shape-Controlled Colloidal Interactions in Nematic Liquid Crystals. \textit{Science} \textbf{326}, 1083-1086 (2009).


\bibitem{Nobili.1992} Nobili M. and Durand G. Disorientation-induced disordering at a nematic-liquid-crystal--solid interface.
\textit{Phys. Rev. A} \textbf{46}, R6174-R6177 (1992).


   \bibitem{Fournier.2005}Fournier J.~B. and  Galatola P. Modeling planar degenerate wetting and anchoring in nematic liquid crystals.
\textit{EPL} \textbf{72}, 403-409 (2005).

   \bibitem{andr} Andrienko D., Allen M. P., Ska\v{c}ej G. and \v{Z}umer S. Defect structures and torque on an elongated colloidal particle immersed in a liquid crystal host. \textit{Phys. Rev. E} \textbf{65}, 041702 (2002).

   \bibitem{hung} Hung F. R. Quadrupolar particles in a nematic liquid crystal: Effects of particle size and shape. \textit{Phys. Rev. E} \textbf{79}, 021705 (2009).

   \bibitem{lapo04} Lapointe C. P., Hultgren A., Silevitch D. M., Felton E. J., Reich D. H. and Leheny R. L. Elastic Torque and the Levitation of Metal Wires by a Nematic Liquid Crystal. \textit{Science}
\textbf{303}, 652-655 (2004).

   \bibitem{copar:13}\v{C}opar, S. and  \v{Z}umer S. Quaternions and hybrid nematic disclinations. {\it Proc. R. Soc. A} {\bf 469}, 20130204 (2013).

   \bibitem{broc} Brochard F. and de Gennes P. G. Theory of Magnetic Suspensions in Liquid Crystals. \textit{J. Phys. (Paris)} \textbf{31}, 691 (1970).

   \bibitem{contpg} The histograms of the angular distributions shown in Fig.~\ref{exp_fig1} were built thanks to a
   standard video tracking program which fits the contour of the ellipsoids' projection in the $xy$-plane. Some
   of the ellipsoids may be slightly tilted out of this plane, but it is neglected in our analysis.

 \bibitem{poul99} Poulin P., Frances N. and Mondain-Monval O. Suspension of spherical particles in nematic solutions of disks and rods. \textit{Phys. Rev. E} \textbf{59}, 4384-4387 (1999).

   \bibitem{lavr98} Lavrentovich O. D.  Topological defects in dispersed liquid crystals, or words and worlds around liquid crystal drops. \textit{Liq. Cryst.} \textbf{24}, 117-125 (1998).

\bibitem{tasinkevych:2002} {Tasinkevych M.,  Silvestre N.~M.,  Patr\'{i}cio P. and Telo da Gama M.~M.  Colloidal interactions in two-dimensional nematics. \textit{Eur. Phys. J. E} {\bf 9}, 341 (2002).}

\bibitem{smalyukh:2005}{Smalyukh I. I.,  Kuzmin A.~N., Kachynski A.~V., Prasad P.~N. and Lavrentovich O. D. Optical trapping of colloidal particles and measurement of the defect line tension and colloidal forces in a thermotropic nematic liquid crystal. \textit{Appl. Phys. Lett.} \textbf{ 86}, 021913 (2005).}


   \bibitem{alex} Alexander G. P., Chen B. G., Matsumoto E. A. and Kamien R. D. Colloquium: Disclination loops, point defects, and all that in nematic liquid crystals \textit{Rev. Mod. Phys.} \textbf{84}, 497-514 (2012).

   \bibitem{quist} Quist P. O., Halle B. and Fur\'{o} I. Micelle size and order in lyotropic nematic phases from nuclear spin relaxation.  \textit{J. Chem. Phys.} \textbf{96}, 3875 (1992).

 \bibitem{nesr} Nesrullajev A.  Shape and sizes of micelles in nematic-calamitic and nematic-discotic mesophases: Sodium lauryl sulphate/water/decanol lyotropic system. \textit{Mater. Chem. Phys.} \textbf{123}, 546-550 (2010).

   \bibitem{pash} Pashley R. M. and Ninham B. W.  Double-layer forces in ionic micellar solutions. \textit{J. Phys. Chem.} \textbf{91}, 2902-2904 (1987).

   \bibitem{monda96} Mondain-Monval O., Leal-Calderon F. and Bibette J. Forces Between Emulsion Droplets: Role of Surface Charges and Excess Surfactant. \textit{J. Phys. II France} \textbf{6}, 1313-1329 (1996).

    \bibitem{poni} Poniewierski A. and Holyst R. Nematic alignment at a solid substrate: The model of hard spherocylinders near a hard wall. \textit{Phys. Rev. A} \textbf{38}, 3721-3727 (1988).

\end{thebibliography}
\end{document}

\bibliographystyle{apsrev}

%\begin{center}{\bf \Large Supplementary Information: Dispersions of ellipsoidal particles in a nematic liquid crystal}\\\vspace{0.5cm}\normalsize
%Mykola Tasinkevych$^{\dag,\ddag,}$\footnote{\url{miko@is.mpg.de}}, Fr\'{e}d\'{e}ric Mondiot$^\S$, Olivier
%Mondain-Monval$^\S$, and Jean-Christophe Loudet$^{\S,}$\footnote{\url{loudet@crpp-bordeaux.cnrs.fr}}\\
%\vspace{0.5cm} $^\dag$\textit{Max-Planck-Institut f\"ur Intelligente Systeme,\\Heisenbergstr. 3, D-70569
%Stuttgart, Germany}\\$^\ddag$\textit{Institut f\"{u}r Theoretische Physik IV,Universit\"{a}t
%Stuttgart,\\Pfaffenwaldring 57, D-70569 Stuttgart, Germany}\\$^\S$\textit{Universit\'{e} de Bordeaux, CNRS,
%Centre de Recherche Paul Pascal, Avenue A. Schweitzer F-33600 Pessac, France}\end{center}\vspace{1cm}

%\begin{center}\Large \textbf{Supplementary Information}\end{center}\normalsize\vspace{1cm}

\title{Supplementary Information: Dispersions of ellipsoidal particles in a nematic liquid crystal}
\author{Mykola Tasinkevych}
\email{miko@is.mpg.de} \affiliation{Max-Planck-Institut f\"ur Intelligente Systeme, Heisenbergstr. 3, D-70569
Stuttgart, Germany} \affiliation{Institut f\"{u}r Theoretische Physik IV,
         Universit\"{a}t Stuttgart, Pfaffenwaldring 57,
         D-70569 Stuttgart, Germany}
\author{Fr\'{e}d\'{e}ric Mondiot}
\affiliation{Universit\'{e} Bordeaux 1, CNRS, Centre de Recherche Paul Pascal, Avenue A. Schweitzer F-33600
Pessac, France}
\author{Olivier Mondain-Monval}
\affiliation{Universit\'{e} Bordeaux 1, CNRS, Centre de Recherche Paul Pascal, Avenue A. Schweitzer F-33600
Pessac, France}
\author{Jean-Christophe Loudet}\email{loudet@crpp-bordeaux.cnrs.fr}
\affiliation{Universit\'{e} Bordeaux 1, CNRS, Centre de Recherche Paul Pascal, Avenue A. Schweitzer F-33600
Pessac, France}

\maketitle

\begin{figure}
\centering
\includegraphics[width=\textwidth]{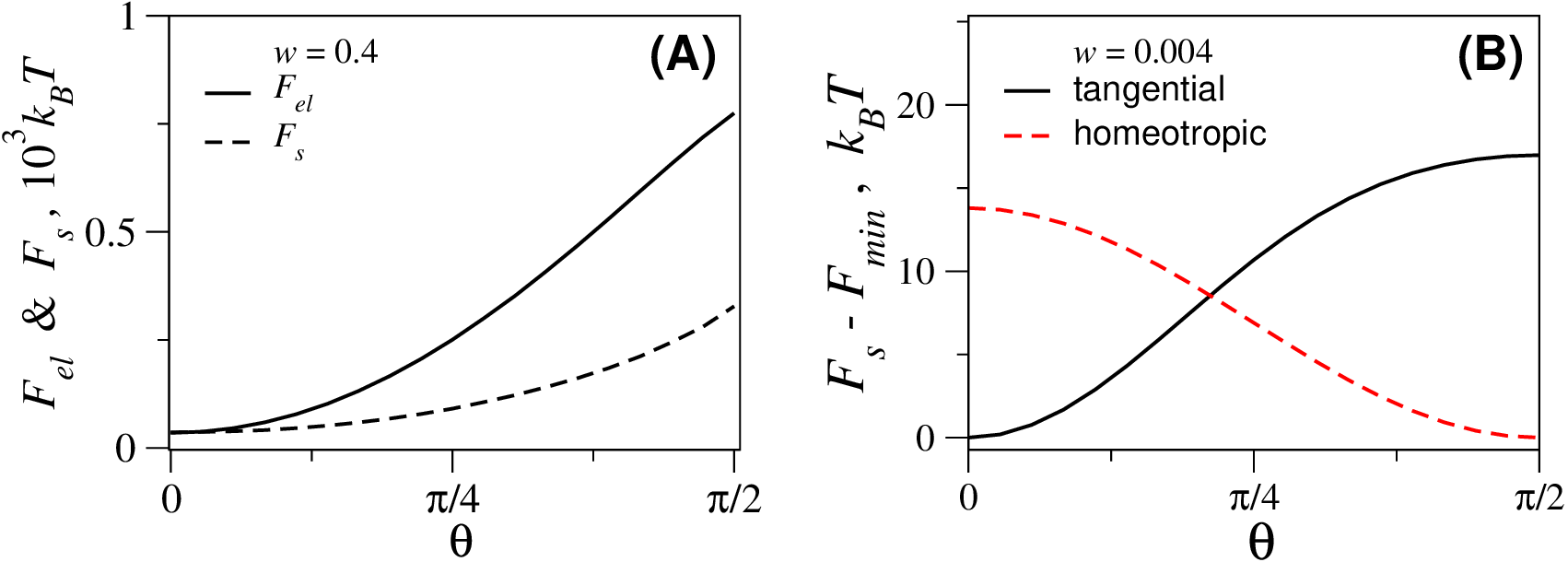}
 \label{ultra_weak_anchoring}
\end{figure}
\begin{center}
\begin{minipage}[c]{\linewidth}
\small{FIG. S1: \textbf{(A)} Elastic ($F_{el}$) and surface anchoring ($F_s$) contributions to the total free energy
 as a function of $\theta$ for weak tangential anchoring ($w = 0.4$) (see main text for details).
\textbf{(B)} Ultra-weak tangential anchoring $w = 0.004\,$, $F_{el} \lesssim 10 ^{-8}$. Aspect ratio $k=5\,$.}
\end{minipage}
\end{center}
\newpage

\noindent {\bf \large Supplementary Note 1 \\ Weak and ultra-weak anchoring} \\
%\noindent {\bf Weak versus ultra-weak anchoring }
%\section{Weak and ultra-weak anchoring}

\noindent  Our main goal here is to provide additional information to explain the parabolic shape of the curves displayed
on Fig.~4 of the main text.\\

\noindent  \textbf{Tangential anchoring}.  As mentioned in the main text, Brochard and de Gennes [54] predicted
that, for a thin rod immersed in a nematic phase with \emph{strong} non-degenerated tangential anchoring, the
elastic free energy $F_{el}$ scales as $\theta^2$, where $\theta$ is the angle between the rod long axis and the
far field director $\dir\,$. The experiments of Lapointe \textit{et al.} confirmed this behavior [52].
The solid black line on Fig.~4 indicates that this $\theta^2$ law seems to hold as well for weak degenerate
anchoring ($w=0.4$). The reason becomes clear if we analyse the contributions from the bulk elastic free energy
$F_{el}$ (the first integral in Eq.~(1) minus the bulk free energy of a uniform nematic) and the surface
anchoring free energy $F_s$ (the second integral in Eq.~(1)), which are both plotted in
Fig.~S1(A) as solid and dashed lines, respectively. We see that, even for $w=0.4\,$, the
bulk nematic distortions are still significant and $F_{el}$ dominates $F_s$ for all $\theta$ values (except at
$\theta=0$ where the two curves join). This result qualitatively explains the quadratic form of the solid black
curve shown in Fig.~4.

Eventually, as the anchoring strength decreases even further, the bulk nematic distortions should fade away and
the total free energy should be determined by the surface anchoring term only. Its contribution can be readily
estimated within the Frank-Oseen model, where the surface free energy may be written in the form $F_s =
\frac{W}{2} \int_{\partial \Omega}({\bf n}  \cdot \bm{\nu})^2 ds\,$. Here, ${\bm{\nu}}$ denotes the surface unit
normal vector and $W>0$ (resp. $W<0$) describes a tangential (resp. homeotropic) anchoring. For $W\ll 1\,$, the
far-field director $\dir$ is only slightly distorted by the presence of a colloidal particle. Assuming $\dir$
along the $z-$axis, the Frank-Oseen elastic free energy (in the one-elastic-constant approximation, $K$) may be
approximated as [54]: $F_{el} \simeq\frac{K}{2}\int d^3r \left [ (\nabla n_x)^2 + (\nabla n_y)^2 \right
]\,$, where $(n_x,n_y,0)$ describes deviations of the local director from $\dir\,$. Consequently, for
$n_{x,y}\ll 1\,$, the leading contribution, $F_{lead}\,$, to the free energy stems from the surface term, i.e.,
$F_{lead} = \frac{W}{2}\int_{\partial \Omega} ({\nu}_z)^2 ds\,$. We may approximate a large-$k$ ellipsoid by a
cylinder with radius $R\,$, then $F_{lead}$ is readily integrated giving
\begin{equation}
   F_{lead} = \pi W k R^2 \left ( \sin^2\theta +\frac{\cos^2\theta}{k} \right )\,,
\label{eq_ultra_low_W}
\end{equation}
where $\theta$ is the angle between the cylinder axis and $\dir\,$. For $k\rightarrow\infty$ (with $kR
\rightarrow \mbox{cst}$), the term $\frac{\cos^2\theta}{k}$, describing the cylinder ends effects, may be
dropped yielding $F_{lead} \propto W \sin^2\theta\,$. Eq.~(\ref{eq_ultra_low_W}) is valid for large $\theta$
too, and it agrees quite well with the numerical results displayed in
Fig.~S1(B), which are obtained for an ultra-weak anchoring strength ($w=0.004$).\\

\noindent  \textbf{Homeotropic anchoring}. It is rather surprising that a similar $\theta^2$ behavior is also observed
for the case of homeotropic anchoring, be it weak (see red dashed curve on Fig.~4) or strong (see open symbols
around the global minima at $\theta =\pi/2$ on Figs.~2 \& 3). We think that sharp deviations from the quadratic
regime, reflected by the onset of a free energy barrier at small $\theta$, and  the rich free energy landscape
are mainly caused by the interplay of the shape anisotropy and the topological defects induced by homeotropic
boundary conditions. \\

%\section{Numerical method}
\noindent {\bf \large Supplementary Note 2 \\ Model parameters and the numerical procedure} \\

\noindent As mentioned in the main text, we used the Landau-de Gennes  theory to compute the total free energy
related to the immersion of prolate ellipsoidal particles in a nematic phase. We here provide additional
information about some variables defined in Eqs.~(1)-(3) of the main text and outline the numerical procedure.

It is convenient to define the dimensionless temperature $\tau=24ac/b^2$. At $\tau<1$ the uniaxial nematic is
stable and the degree of orientational order is given by
\begin{equation}
Q_b=\frac{b}{8c}\left(1+\sqrt{1-\frac{8\tau}{9}}\right)\,.
\end{equation}
The nematic becomes unstable at $\tau>9/8$. At $\tau=1$ both the nematic and the isotropic phases coexist. We
use the following values of the model parameters: $a_0=0.044\times10^6$ J $/$Km$^3$, $b=0.816\times10^6$
J$/$m$^3$, and $c=0.45\times10^6$ J$/$m$^3$, $T^* = 307$ K, $L_1=6\times10^{-12}$J$/$m (these are typical values
for 5CB \cite{kralj.1991}), and for simplicity we put $L_2= 0\,$. The spatial extension of inhomogeneous regions
and the cores of topological defects is of the order of the bulk correlation length, $\xi\,$, given by
$\xi=\left(8c\left(3L_1+2L_2\right)/b^2\right)^{1/2}\simeq10\mathrm{nm}$ at the nematic-isotropic (NI)
transition \cite{Chandresakhar.1992}. The LdG elastic constants $L_1$ and $L_2$ may be related to the FO elastic
constants, $K_1=K_3$ and $K_2$, through the uniaxial ansatz $\Qij=3 Q_b\left(n_i n_j -\delta_{ij}/3 \right)/2$,
yielding $K_{1} = K_{3} = 9 Q_b^2(L_1+ L_2/2)/2$ and $K_{2} = 9Q_b^2L_1/2$ \cite{Frank.1958}. In general, $K_1$
and $K_3$ are different, but in most cases the difference is small and the LdG free energy is deemed adequate.

We minimize the Landau-de Gennes free energy (Eq.~(1), main text) numerically by using adaptive finite elements
methods. Initially, the surface of a colloidal particle is triangulated using Open Source GNU Triangulated
Surface (GTS) library \cite{GTS}. Then, the triangulation of the nematic domain $\Omega$ is carried out using
\textit{Quality Tetrahedral Mesh Generator} \cite{tetgen}, which supports the adaptive mesh refinement. Linear
triangular and tetrahedral elements are used in $2D$ and $3D$, respectively. Generalized Gaussian  quadrature
rules for multiple integrals \cite{cub_encycl} are used in order to evaluate integrals over elements. In
particular, for tetrahedra a fully symmetric cubature rule with 11 points \cite{keast1986} is used, and
integrations over triangles are done by using a fully symmetric quadrature rule with 7 points \cite{stroud1971}.
The discretized Landau-de Gennes functional in then minimized using the National Institute for Research in
Computer Science and Control \textit{M1QN3} \cite{M1QN3} optimization routine, which  implements a limited
memory quasi-Newton technique of Nocedal \cite{nocedal1980}. More details on the numerical implementation can be
found  in ref.~\cite{tasinkevych:2012}.\\

We only consider cylindrically symmetric ellipsoidal particles with semi-axes $(A,B,B)\,$, where $A$ (resp. $B$)
is the semi-long (resp. semi-short) axis. Particles with aspect ratio $k\equiv \frac{A}{B} = 3,5,7\,$, and with
a fixed volume $V=\frac{4}{3}\pi R_0^3$ are analyzed. In the single particle case, we use $R_0 = 0.3\,\mu$m. In
the case of two ellipsoids, we study only the case $k=4\,$, and set $B=0.1\,\mu$m. \\

\noindent {\bf \large Supplementary References}